\documentclass[english,draftcls, onecolumn,12pt]{IEEEtran}
\usepackage[T1]{fontenc}
\usepackage[latin9]{inputenc}
\usepackage{float}
\usepackage{textcomp}
\usepackage{amsthm}
\usepackage{amsmath}
\usepackage{amssymb}
\usepackage{graphicx}
\usepackage{esint}

\makeatletter

\providecommand{\tabularnewline}{\\}
\floatstyle{ruled}
\newfloat{algorithm}{tbp}{loa}
\providecommand{\algorithmname}{Algorithm}
\floatname{algorithm}{\protect\algorithmname}

  \theoremstyle{plain}
  \newtheorem{thm}{\protect\theoremname}
 \ifx\proof\undefined\
   \newenvironment{proof}[1][\proofname]{\par
     \normalfont\topsep6\p@\@plus6\p@\relax
     \trivlist
     \itemindent\parindent
     \item[\hskip\labelsep
           \scshape
       #1]\ignorespaces
   }{%
     \endtrivlist\@endpefalse
   }
   \providecommand{\proofname}{Proof}
 \fi
  \theoremstyle{remark}
  \newtheorem{rem}{\protect\remarkname}
  \theoremstyle{plain}
  \newtheorem{cor}{\protect\corollaryname}
  \theoremstyle{plain}
  \newtheorem{lem}{\protect\lemmaname}

\makeatother

\usepackage{babel}
\providecommand{\corollaryname}{Corollary}
\providecommand{\lemmaname}{Lemma}
\providecommand{\remarkname}{Remark}
\providecommand{\theoremname}{Theorem}

\begin{document}

\title{Optimal Scheduling and Power Allocation for Two-Hop Energy Harvesting
Communication Systems }

\author{Yaming Luo, Jun Zhang, and Khaled B. Letaief, \textit{Fellow, IEEE}%
\thanks{The authors are with the Dept. of Electronic and Computer Engineering,
Hong Kong University of Science and Technology, Hong Kong. Email:
\{luoymhk, eejzhang, eekhaled\}@ust.hk.%
}}
\maketitle
\begin{abstract}
Energy harvesting (EH) has recently emerged as a promising technique
for green communications. To realize its potential, communication
protocols need to be redesigned to combat the randomness of the harvested
energy. In this paper, we investigate how to apply relaying to improve
the short-term performance of EH communication systems. With an EH
source and a non-EH half-duplex relay, we consider two different design
objectives: 1) short-term throughput maximization; and 2) transmission
completion time minimization. Both problems are joint scheduling and
power allocation problems, rendered quite challenging by the half-duplex
constraint at the relay. A key finding is that directional water-filling
(DWF), which is the optimal power allocation algorithm for the single-hop
EH system, can serve as guideline for the design of two-hop communication
systems, as it not only determines the value of the optimal performance,
but also forms the basis to derive optimal solutions for both design
problems. Based on a relaxed energy profile along with the DWF algorithm,
we derive key properties of the optimal solutions for both problems
and thereafter propose efficient algorithms. Simulation results will
show that both scheduling and power allocation optimizations are necessary
in two-hop EH communication systems.\end{abstract}

\section{Introduction}

The growing concerns on the energy consumption of wireless networks
and its associated global warming effects have spurred lots of research
activities related to the development of more energy-efficient communication
techniques. Energy harvesting (EH) has recently emerged as a promising
approach to realize green communications, as it can power the communication
devices and networks with renewable energy sources. Communication
terminals with EH capability can harvest energy from the environment
\cite{EH_WSN}, including solar energy, vibration energy, thermoelectric
energy, RF energy, etc. However, as the harvested energy is typically
in a small amount and also random, how to guarantee satisfactory short-term
performance is a challenging problem.

\subsection{Related Works and Motivations}

The potential of EH technology has recently spurred lots of research
activities in the wireless communications community. The capacity
of a point-to-point link with an EH transmitter was investigated in
\cite{EH_info1} for the AWGN channel and in \cite{EH_info2} for
the fading channel, respectively. To achieve the capacity, new transmission
policies such as save-and-transmit and best-effort-transmit schemes
are required. In addition to these information-theoretic studies,
practical transmission policies have also been investigated. An offl{}ine
packet scheduling problem to minimize the transmission completion
time with a discrete EH model and infi{}nite energy storage at the
transmitter was fi{}rst introduced in \cite{EH_DWF_time}, and this
work was later extended in \cite{EH_DWF_battery} to transmitters
with limited energy storage. A directional water-fi{}lling algorithm
adapted to the instantaneous channel state over a Gaussian fading
channel was developed in \cite{EH_DWF_channel}, while channel training
optimization was investigated in \cite{EH_train}. Other EH communication
systems were also investigated, including the broadcast channel \cite{EH_DWF_Broad,EH_Broad_time},
the multiple access channel \cite{EH_DWF_Mac}, and the interference
channel \cite{EH_interference}. These studies uncovered an important
and unique property of EH communication systems; namely, even when
the channel remains unchanged, the transmit power should adapt to
the random energy arrivals. Subsequently, communication protocols
need to be revised in EH systems for satisfactory short-term performance.

Multi-hop transmission is often adopted to increase the communication
range and utilize the energy more efficiently in wireless communication
systems. Hence, it can serve as a potential candidate to enhance the
short-term performance of EH communication networks. However, relaying
protocols need to be re-designed for the random energy arrival. Therefore,
further investigation is required in order to exploit the potential
of multi-hop transmissions. Energy harvesting two-hop networks have
been studied in \cite{EH_two_hop1,EH_two_hop2,EH_two_hop3}, but only
for some special cases. Specifically, the optimal transmission policy
with a non-EH source and an EH relay was developed in \cite{EH_two_hop1},
while the case of an EH source with two energy arrivals was considered
in \cite{EH_two_hop2} and \cite{EH_two_hop3}. Moreover, these previous
works only considered the throughput as the objective.

So far, the optimal transmission policy for a two-hop EH communication
system with a general EH source of multiple energy arrivals is still
not known. This is in fact a more important case as the EH source
has a larger effect on the design and performance of the whole system
than the EH relay. Moreover, the scheduling with a general EH source
is more challenging. For example, if only the relay is an EH node,
the optimal scheduling involves a two-stage transmission \cite{EH_two_hop1},
but as will be shown in this paper, generally, multi-stage scheduling
is needed when the source is an EH node. As power adaptation is also
needed at the EH source to combat the random energy arrivals, we are
faced with a joint scheduling and power allocation problem.

\subsection{Contributions}

In this paper, we will investigate a two-hop communication system
with an EH source and a non-EH relay. Two different objectives will
be considered. The first is a short-term throughput maximization problem
with a given deadline, while the second is a transmission completion
time minimization problem with a given amount of data. We find that
the optimal power allocation algorithm for the single-hop EH transmission,
namely, directional water-filling (DWF), can serve as guideline for
the design of the two-hop system. Essentially, the DWF power allocation
results for the single-hop transmission will help us decouple the
joint scheduling and power allocation problem. Specifically, for any
given EH profile at the source, we can first find a related DWF EH
profile, which provides a performance upper bound and the optimal
transmission policy for which can be derived. Then, we can extend
the result to solve the original design problem. Efficient algorithms
are then proposed to achieve the performance upper bound, and guarantee
optimality. Simulation results shall demonstrate the importance of
the adaptive power allocation at the EH source and the optimal scheduling
of the source and relay transmission periods. Particularly, both of
the fixed power allocation and the fixed scheduling transmission policies
always incur a performance loss with a time-varying EH profile, especially
when there is a significant difference between the energy levels of
the source and relay. 

The organization of this paper is as follows. In Section II, we introduce
the energy harvesting model and formulate two design problems: Short-term
throughput maximization, and transmission completion time minimization.
In Sections III and IV, the two problems are solved with the help
of the DWF EH profile. Simulation results are given in Section V.
Finally, Section VI summarizes our work.

\section{System Model and Problem Formulation}

We consider a two-hop communication system with a half-duplex decode-and-forward
relay, where the direct link between the source and destination is
too weak to provide reliable communication. The source is an EH node,
all the energy harvested by which is used for communication. The relay
is a non-EH node, powered by a battery, to assist the communication
between the EH source and its destination. For the data transmission
of both the source-relay (S-R) and relay-destination (R-D) channels,
we assume that the power-rate relationship follows a non-negative,
strictly concave and monotonically increasing function, $R=g(P)$,
where $P$ is the instantaneous receive power. These properties are
satisfied in many widely used communication models. For the relay
transmission, there exists a data causality constraint, which means
that at time instant $t$, the data transmitted by the relay, denoted
as $D^{\textrm{R}}\left(t\right)$, should not exceed that transmitted
by the source, denoted as $D^{\textrm{S}}\left(t\right)$. For ease
of reference, we list the main notations defined in this section in
Table I.

\begin{table}
\caption{Main notations}

\centering{}%
\begin{tabular}{|c||l|}
\hline 
Symbols & Definition\tabularnewline
\hline 
\hline 
$T$ & Total transmission period\tabularnewline
\hline 
$N$ & Number of energy arrivals in $T$\tabularnewline
\hline 
$P^{\textrm{S}}\left(t\right)$/$P^{\textrm{R}}\left(t\right)$ & Instantaneous transmit power of the source/relay\tabularnewline
\hline 
$D^{\textrm{S}}\left(t\right)$/$D^{\textrm{R}}\left(t\right)$ & Amount of data transmitted until $t$ by the source/relay\tabularnewline
\hline 
$E_{\Sigma}^{\textrm{EH}}\left(t\right)$ & Cumulative harvested energy at time $t$\tabularnewline
\hline 
$t_{k}$ & $k$-th energy arrival epoch\tabularnewline
\hline 
$E_{k}$ & Energy harvested in the $k$-th energy arrival interval\tabularnewline
\hline 
$E_{\Sigma}$ & Total harvested energy in $T$\tabularnewline
\hline 
$N^{\textrm{S}}$ & Number of all the source-relay stage pairs\tabularnewline
\hline 
$P_{M}^{\textrm{R}}$ & Maximum transmit power of the relay\tabularnewline
\hline 
$E_{\Sigma}^{\textrm{R}}$ & Total amount of energy available at the relay\tabularnewline
\hline 
\end{tabular}
\end{table}

\subsection{Energy Harvesting Model}

\begin{figure}
\begin{centering}
\includegraphics[scale=0.65]{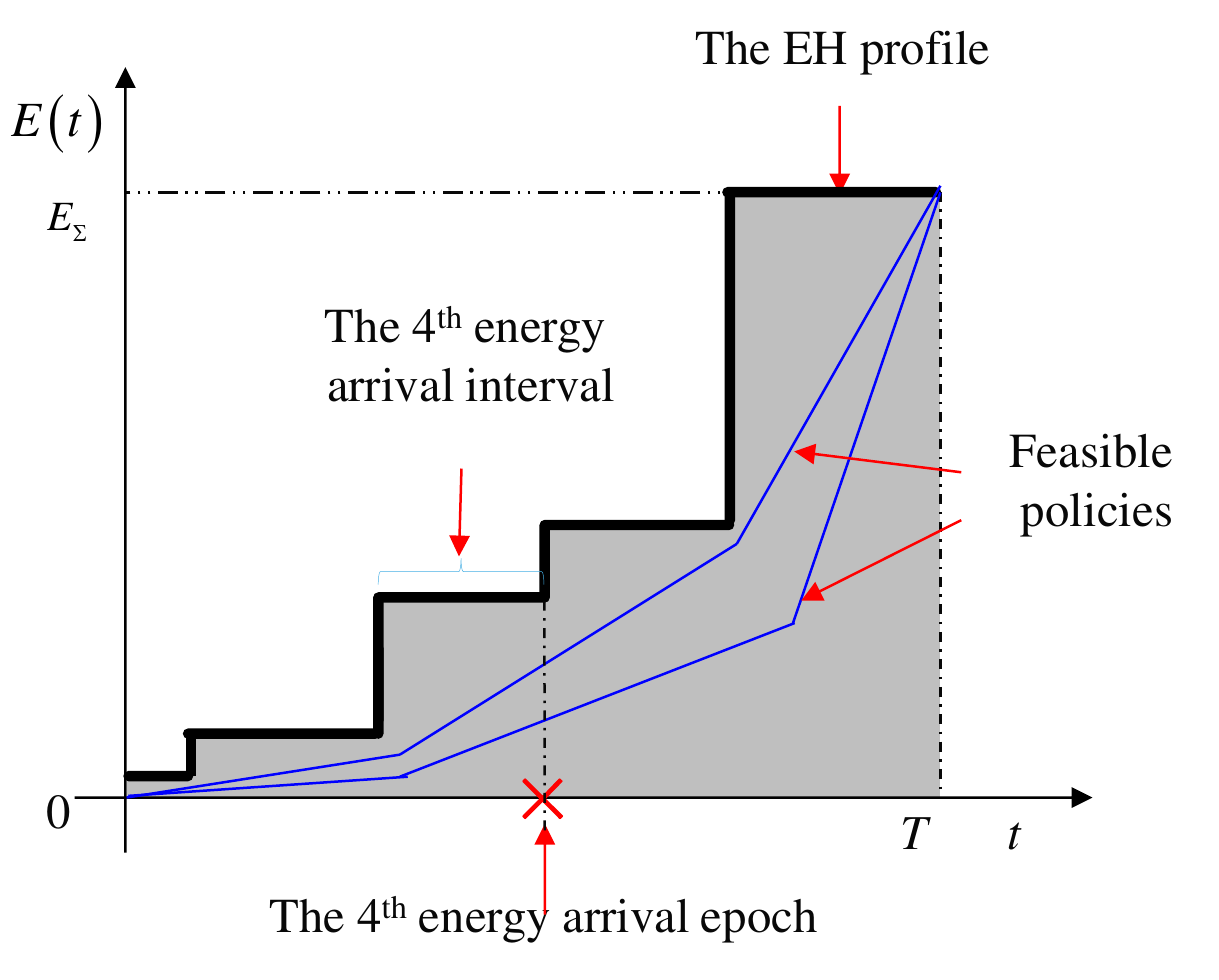}
\par\end{centering}

\caption{\label{fig:An-example-of}An example of the staircase-shaped EH profile.
The power consumption curves under the EH profile denote feasible
policies, for which the slope represents the instantaneous transmit
power. }
\end{figure}

An important factor that determines the performance of an EH system
is the \textit{EH profile}, denoted as $E_{\Sigma}^{\textrm{EH}}\left(t\right)$,
which models the cumulative harvested energy up to time $t$. Similar
to \cite{EH_DWF_time,EH_DWF_channel,EH_two_hop1,EH_two_hop2,EH_two_hop3},
we consider a discrete-time EH profile, as shown in Fig. \ref{fig:An-example-of}.
Such discrete-time EH profiles can also be used to approximate other
EH profiles, and later we will show that our results can deal with
more general EH profiles. To demonstrate the impact of EH profiles,
we assume that the information of the EH profile in the coming transmission
block is available at the beginning of the block, which is also adopted
in the previously-mentioned works. We refer to the time instant when
the energy arrives as the \emph{energy arrival epoch}, denoted as
$t_{i}$, $i=1,2,...,N$, with $t_{N}<T$, where $T$ is the total
transmission period, and $N$ is the number of energy arrivals in
the whole transmission period. The corresponding amount of harvested
energy in the $i$-th arrival is denoted as $E_{i}$, and the respective
cumulative energy is $E_{\Sigma}^{\textrm{EH}}\left(t_{i}\right)\triangleq\sum_{j=1}^{i}E_{j}$.
The time interval between the $(i-1)$-th and the $i$-th energy arrival
epochs is called the $i$-th \emph{energy arrival interval}. For convenience,
the interval between the $N$-th energy arrival and the ending time
$T$ is called the $\left(N+1\right)$-th energy arrival interval.
Without loss of generality, we assume that $t_{1}=0$, and that the
total harvested energy is $E_{\Sigma}$. 

The utilization of the harvested energy is constrained by the EH profile,
which yields the causal energy neutrality constraint \cite{Power_EH},
i.e., the energy consumed thus far cannot exceed the total harvested
energy. Denote the transmit power at time $t$ as $P\left(t\right)$,
then the energy neutrality constraint can be expressed as 
\begin{equation}
\int_{0}^{t}{P}\left(\tau\right)d\tau\le E_{\Sigma}^{\textrm{EH}}\left(t\right).\label{eq:1}
\end{equation}
Accordingly, a certain EH profile determines a feasible energy consumption
domain, and only the transmission policies inside this domain are
feasible, as shown in Fig. \ref{fig:An-example-of}. Due to this energy
neutrality constraint, we cannot use the energy arriving in the future,
but can save the current energy for future use. 

Besides the EH profile, the battery capacity is also an important
factor for the EH link performance. In this paper, we assume that
the battery capacity is infinite, while the case with a finite capacity
will be considered in future work.

\subsection{Transmission Policy}

Denote the instantaneous source (or relay) transmit power as $P^{\textrm{S}}\left(t\right)$
(or $P^{\textrm{R}}\left(t\right)$). With a random EH profile, transmitting
with a constant power will in general not be optimal and the transmit
power needs to be adjusted according to the EH profile. Meanwhile,
due to the half-duplex constraint at the relay, i.e., $P^{\textrm{S}}\left(t\right)P^{\textrm{R}}\left(t\right)=0,$
$\forall t$, the whole transmission block will be divided into multiple
\emph{stages}, in each of which only one side (source or relay) is
allowed to transmit. The stage in which the source (relay) transmits
is called the \emph{source (relay) stage}, and the union of all the
source (relay) stages in the $k$-th DWF interval is denoted as $\mathfrak{T}_{k}^{\textrm{S}}$
($\mathfrak{T}_{k}^{\textrm{R}}$). The length of each source (relay)
stage is called a \emph{source (relay)} \emph{period}%
\footnote{In this paper, we make an ideal assumption that each source or relay
stage can take any real value. In practice, there will be a minimum
time unit related to the symbol duration, and our results can be round
off to serve as approximations. A detailed investigation of this aspect
is left to future work.%
}, and then \emph{scheduling} means the transmission time allocation
between the source and relay. Therefore, the design of transmission
policies in two-hop EH communication systems involves power allocation
at the EH source and scheduling between the source and relay transmissions. 

\begin{figure}
\begin{centering}
\includegraphics[scale=0.72]{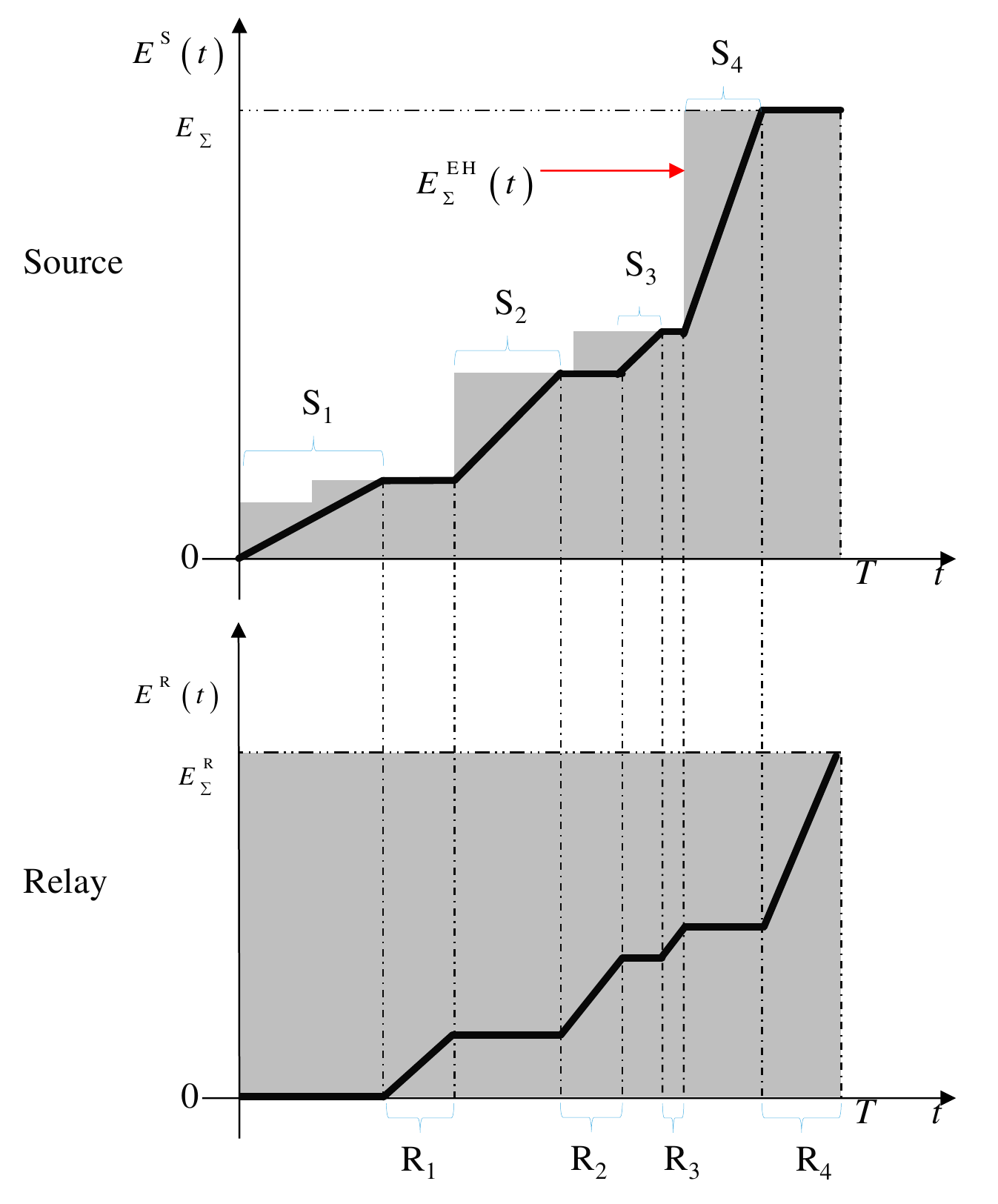}
\par\end{centering}

\caption{\label{fig: power consumption curves }An example of the source/relay
scheduling and power allocation, represented by the bold solid curves.
On one hand, the slope of the curve represents the instantaneous transmit
power. On the other hand, at each time instant, the nonzero source/relay
transmit power signifies a source/relay stage. In this example, $\textrm{S}_{i}$
($i=1,2,3,4$) is a source stage, while $\textrm{R}_{i}$ ($i=1,2,3,4$)
is a relay stage. }
\end{figure}

An example of the source/relay scheduling and power allocation is
illustrated in Fig. \ref{fig: power consumption curves }, the slope
of which represents the instantaneous transmit power. At each time
instant, if the source/relay transmit power is nonzero, it belongs
to a source/relay stage. With the half-duplex constraint, either the
source or the relay power consumption curve completely describes a
scheduling result. Therefore, we only adopt the source power consumption
curve to represent a scheduling result in the rest of this paper.
Due to the data causality constraint, the last stage must be a relay
stage, while the first stage must be a source stage. The total number
of stages should be an even number, and source and relay will take
turns to transmit, but the length of each stage will in general be
different from each other, as shown in Fig. \ref{fig: power consumption curves }.
Each source stage in combination with the relay stage directly after
it is called an \emph{S-R stage pair.} The total number of S-R stage
pairs is denoted as $N^{\textrm{S}}$. 

For the case of a non-EH source and an EH relay, which was tackled
in \cite{EH_two_hop1}, the relay tends to wait for enough energy,
and delaying the relay transmission can always get a better performance
without violating either the data or energy feasibility condition.
Hence, the optimal scheduling is to divide the whole transmission
block into two stages, i.e., a single S-R stage pair. For the system
with a non-EH relay and an EH source as in our case, it is the source
that wants to wait for enough energy. However, the two-stage transmission
will in general be sub-optimal due to the data causality constraint.
Therefore, the optimal scheduling is more challenging with an EH source
than with an EH relay.

\subsection{Problem Formulation }

Since the short-term performance of EH systems is more crucial, in
the context of two-hop communications, we consider two different objectives:
Short-term throughput maximization and transmission completion time
minimization. For the first objective, the total transmission period
is fixed, while for the second, the total amount of data that needs
to be transmitted is fixed. To make the problem tractable, we assume
that both the S-R and R-D channels are static. For the non-EH relay,
we consider two kinds of constraints \cite{Power_constraint}: 1)
a peak power constraint, which means that the instantaneous transmit
power should not exceed a threshold $P_{M}^{\textrm{R}}$, due to
the linear operation of power amplifiers or safety regulation; and
2) a total energy constraint, which means that the total consumed
energy should not exceed the total available energy $E_{\Sigma}^{\textrm{R}}$,
the capacity of the battery. 

Given the above, the short-term throughput maximization problem (denoted
as RMAX) can be formulated as follows:

\noindent 
\begin{align}
\begin{array}{ccc}
\textbf{RMAX:}\end{array}\underset{P{}^{\textrm{S}}\left(t\right),P{}^{\textrm{R}}\left(t\right)}{\max} & \phantom{=}D{}^{\textrm{R}}\left(T\right) & \phantom{=}\label{eq:2}\\
\textrm{\ensuremath{\phantom{==}}s.t.\ensuremath{\phantom{==}}} & P{}^{\textrm{S}}\left(t\right)\geq0,\begin{array}{c}
\end{array}P{}^{\textrm{R}}\left(t\right)\geq0,\begin{array}{c}
\end{array}P{}^{\textrm{S}}\left(t\right)P{}^{\textrm{R}}\left(t\right)=0, & \phantom{=}\label{eq:3}\\
 & \int_{0}^{t}P{}^{\textrm{S}}\left(\tau\right)d\tau\leq E_{\Sigma}^{\textrm{EH}}\left(t\right),\begin{array}{c}
\end{array}D{}^{\textrm{R}}\left(t\right)\leq D{}^{\textrm{S}}\left(t\right), & \phantom{=}\label{eq:4}\\
 & \int_{0}^{T}P{}^{\textrm{R}}\left(\tau\right)d\tau\leq E_{\Sigma}^{\textrm{R}},\begin{array}{c}
\end{array}P{}^{\textrm{R}}\left(t\right)\leq P_{M}^{\textrm{R}},\begin{array}{c}
\end{array}t\in\left[0,T\right], & \phantom{=}\label{eq:5}
\end{align}
where (\ref{eq:3}) is due to the half duplex constraint at the relay,
(\ref{eq:4}) is due to the energy causality constraint and the data
causality constraint, (\ref{eq:5}) is due to the total energy constraint
and the peak power constraint at the relay, while the objective function
(\ref{eq:2}) is derived as $\textrm{min}\left\{ D{}^{\textrm{S}}\left(t\right),D{}^{\textrm{R}}\left(T\right)\right\} =D{}^{\textrm{R}}\left(T\right)$
due to (\ref{eq:4}).

Similarly, the transmission completion time minimization problem (denoted
as TMIN) is as follows:

\noindent 
\begin{align*}
\begin{array}{ccc}
\textbf{ TMIN:}\end{array}\underset{P{}^{\textrm{S}}\left(t\right),P{}^{\textrm{R}}\left(t\right)}{\min} & T\\
\textrm{s.t.\ensuremath{\phantom{==}}} & D{}^{\textrm{R}}\left(T\right)=D,\\
 & P{}^{\textrm{S}}\left(t\right)\geq0,\begin{array}{c}
\end{array}P{}^{\textrm{R}}\left(t\right)\geq0,\begin{array}{c}
\end{array}P{}^{\textrm{S}}\left(t\right)P{}^{\textrm{R}}\left(t\right)=0,\\
 & \int_{0}^{t}P{}^{\textrm{S}}\left(\tau\right)d\tau\leq E_{\Sigma}^{\textrm{EH}}\left(t\right),\begin{array}{c}
\end{array}D{}^{\textrm{R}}\left(t\right)\leq D{}^{\textrm{S}}\left(t\right),\\
 & \int_{0}^{T}P{}^{\textrm{R}}\left(\tau\right)d\tau\leq E_{\Sigma}^{\textrm{R}},\begin{array}{c}
\end{array}P{}^{\textrm{R}}\left(t\right)\leq P_{M}^{\textrm{R}},\begin{array}{c}
\end{array}t\in\left[0,T\right].
\end{align*}

For both RMAX and TMIN problems, due to the half-duplex constraint
$P{}^{\textrm{S}}\left(t\right)P{}^{\textrm{R}}\left(t\right)=0$,
the scheduling and power allocation are coupled. For a given solution,
once we modify the scheduling decision, the power allocation result
should also be adjusted, and vice versa. Moreover, compared to the
two-hop system with a non-EH source in \cite{EH_two_hop1}, our problem
is more challenging, as with an EH source, there will be multiple
S-R stage pairs, depending on the EH profile. Therefore, a more complicated
scheduling is needed. Since it is difficult to solve the optimization
problems directly, we will seek an indirect approach to get the optimal
solutions.

\section{Short-term Throughput Maximization }

In this section, we will solve the RMAX problem. We will first show
that a kind of DWF EH profile that is based on the original EH profile
can provide a throughput upper bound. We will then verify that this
upper bound can be achieved by the original EH profile, and efficient
algorithms will then be proposed to solve the RMAX problem.

\subsection{The DWF EH profile }

\begin{figure}
\begin{centering}
\includegraphics[scale=0.65]{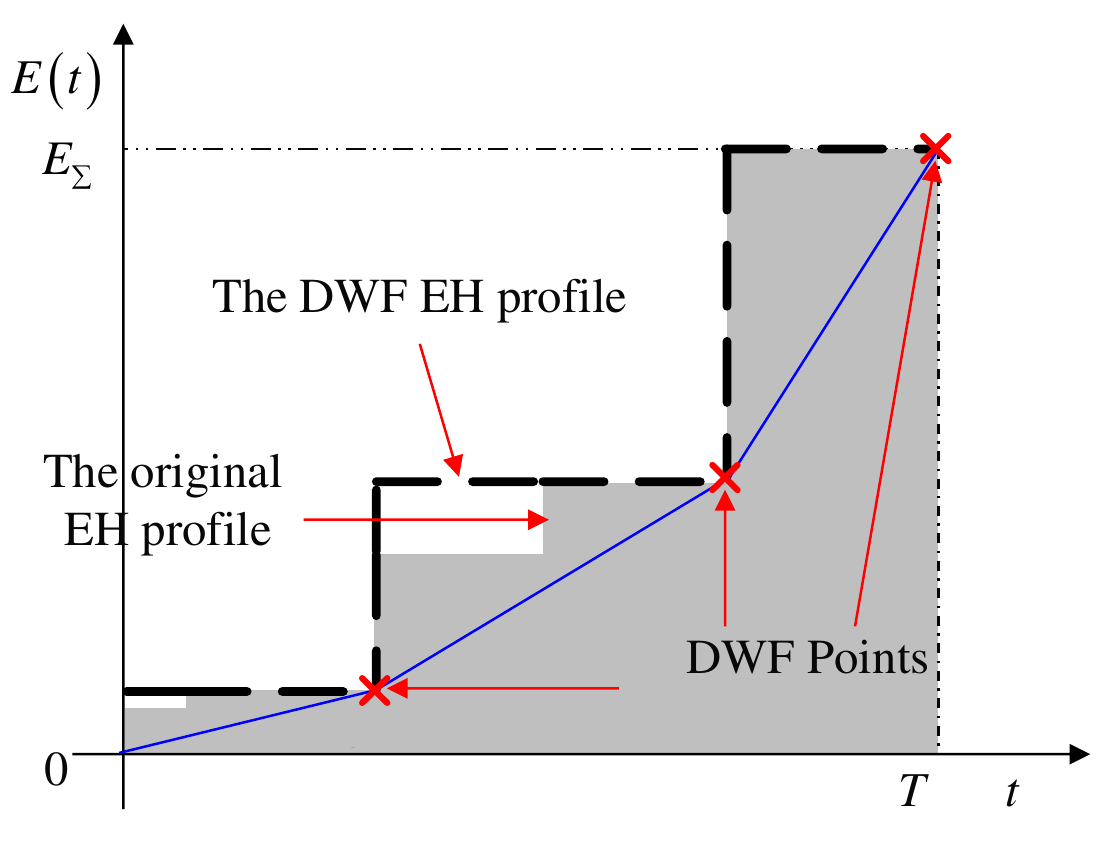}
\par\end{centering}

\caption{\label{fig:The-DWF-power}The DWF power allocation for a given EH
profile, and its associated DWF EH profile. The edge of the dark area
denotes the original EH profile. The DWF points are marked by crosses,
while the curve that connects the DWF points is the DWF power allocation
curve for the single-hop transmission, the slope of which represents
the transmit power. }
\end{figure}

In the single-hop EH communication system with random energy arrivals,
as shown in Fig. \ref{fig:The-DWF-power}, the optimal transmit power
allocation yields a directional water-filling interpretation \cite{EH_DWF_channel,DWF_all},
which has the following three properties: 
\begin{enumerate}
\item The transmit power increases monotonically.
\item The transmit power remains constant between the energy arrivals. 
\item Whenever the power level changes, the energy consumed up to that time
instant equals the total harvested energy.
\end{enumerate}
According to these properties, the transmit power can only change
at some of the energy arrival epochs, and up to these time instants
all the energy harvested should be exhausted. We define those points
where the power level changes as \emph{DWF points}, denoted as $\left(t_{k}^{\textrm{D}},\sum_{i=1}^{k}E{}_{i}^{\textrm{D}}\right),$
$1\leq k\leq N{}^{\textrm{D}}$, where $N{}^{\textrm{D}}$ represents
the number of DWF points, $t_{k}^{\textrm{D}}$ is the horizontal
coordinate of the DWF point, and $E{}_{i}^{\textrm{D}}$ is the amount
of the energy harvested between the $\left(i-1\right)$-th and $i$-th
DWF points. Particularly, the DWF points exclude the origin but include
the ending point. The DWF points can be determined as $t{}_{k}^{\textrm{D}}=t_{i_{k}}$,
$E_{k}^{\textrm{D}}=\sum_{j=i_{k-1}+1}^{i_{k}}E_{j}$, where all the
$i_{k}$ can be computed iteratively as 
\begin{equation}
i_{k}=\arg\underset{i:t_{i}\leq T}{\min}\left\{ \frac{\sum_{j=i_{k-1}+1}^{i}E_{j}}{t{}_{i}-t_{i_{k-1}}}\right\} \label{eq:6}
\end{equation}
with $i_{0}=1$ and $t_{0}=0$. We also define the interval between
two adjacent DWF points (or the origin) as a \emph{DWF interval},
and the length of the $k$-th DWF interval is denoted as $T_{k}^{\textrm{D}}$.

Throughout the paper, we will demonstrate that these DWF points obtained
in the single-hop case are extremely important for the optimization
of the two-hop case. For two-hop communications, according to these
DWF points, we can construct a \emph{DWF EH profile}, which only includes
the beginning time instants of these DWF intervals as its energy arrival
epochs, as illustrated in Fig. \ref{fig:The-DWF-power}. Since this
DWF EH profile is relaxed from the original EH profile and has a larger
feasible domain, it can provide a performance upper bound for the
original EH profile. The DWF EH profile can be uniquely determined
by the set of all the DWF points, and vice versa. For ease of reference,
we list the main notations related to the DWF EH profile in Table
II.

\begin{table}
\caption{Main notations related to The DWF EH Profile}

\centering{}%
\begin{tabular}{|c||l|}
\hline 
Symbols & Definition\tabularnewline
\hline 
\hline 
$i_{k}$ & Index of the $k$-th DWF point in all the energy arrival epochs\tabularnewline
\hline 
$t_{k}^{\textrm{D}}$ & Time instant of the the $k$-th DWF point\tabularnewline
\hline 
$T_{k}^{\textrm{D}}$ & Time duration of the $k$-th DWF interval\tabularnewline
\hline 
$E{}_{k}^{\textrm{D}}$ & Total harvested energy in the $k$-th DWF interval\tabularnewline
\hline 
$N^{\textrm{D}}$ & Number of all the DWF intervals\tabularnewline
\hline 
\end{tabular}
\end{table}

\subsection{Throughput Upper Bound from the DWF EH profile}

We will first analyze the properties of the optimal transmission policy
for an EH source with the DWF EH profile, and then obtain the respective
optimal solution based on these properties. 

Note that the optimal transmission policy is not unique, i.e., different
scheduling decisions may provide the same performance. Unless otherwise
mentioned, if the problem has multiple optimal solutions, we select
the one with the smallest number of S-R stage pairs. For the DWF EH
profile, since all of the energy in each DWF interval is available
at the beginning of that DWF interval, the number of S-R stage pairs
inside each DWF interval can always attain the minimum value as 1.
The properties of the optimal solution to RMAX are listed in Theorem
\ref{thm:Property}. 
\begin{thm}
\label{thm:Property}For \textbf{Problem} \textbf{RMAX} with a source
having the DWF EH profile, the optimal transmission policy has the
following properties:

1) Within the $k$-th DWF interval,\textup{ $k=1,2,...,N^{\textrm{D}}$},
\textup{
\[
\begin{array}{cc}
P^{{\rm S}}\left(t\right)=\Biggl\{\begin{array}{cc}
P_{k}^{{\rm {S}}} & t\in\mathfrak{T}_{k}^{\textrm{S}}\\
0 & t\in\mathfrak{T}_{k}^{\textrm{R}}
\end{array}, & P^{{\rm R}}\left(t\right)=\Biggl\{\begin{array}{cc}
P_{k}^{{\rm {R}}} & t\in\mathfrak{T}_{k}^{\textrm{R}}\\
0 & t\in\mathfrak{T}_{k}^{\textrm{S}}
\end{array},\end{array}
\]
}

where both\textup{ $P_{k}^{\textrm{S}}$} and \textup{$P_{k}^{\textrm{R}}$}
are constants to be determined.

2) At the end of the $k$-th DWF interval, \textup{$k=1,2,...,N^{\textrm{D}}$},\textup{
\[
\begin{array}{cc}
\int_{0}^{t_{k}^{\textrm{D}}}P{}^{\textrm{S}}\left(\tau\right)d\tau=E_{\Sigma}^{\textrm{EH}}\left(t_{k}^{\textrm{D}}\right), & D^{\textrm{R}}\left(t_{k}^{\textrm{D}}\right)=D^{\textrm{S}}\left(t_{k}^{\textrm{D}}\right).\end{array}
\]
}\end{thm}
\begin{IEEEproof}
The proof is given in Appendix 1.\end{IEEEproof}
\begin{rem}
The physical meaning of Property 1 is that the source (relay) should
use the same transmit power in all the source (relay) stages within
a given DWF interval. Property 2 means that at the end of each DWF
interval, the relay empties its data buffer (denoted as the relay
data equality); while the source empties its energy buffer (denoted
as the source energy equality). Also note that Theorem \ref{thm:Property}
holds for both the energy and power constraints at the relay.
\end{rem}
Next we will derive the optimal solution for Problem RMAX based on
the properties in Theorem \ref{thm:Property}. We first consider an
energy constraint at the relay and then extend the result to the general
case with both power and energy constraints. Denote the amount of
relay energy consumption in the $k$-th DWF interval as $E_{k}^{\textrm{R}}$,
and the $k$-th source period as $T_{k}^{\textrm{S}}$. The optimal
solution for Problem RMAX with an energy constrained relay is provided
in the following Corollary. 
\begin{cor}
\label{cor:Solution energy}In the optimal transmission policy for\textbf{
Problem} \textbf{RMAX} with the DWF EH profile and an energy constrained
relay, there is a single S-R stage pair in each DWF interval, i.e.,
the source and relay stages for the $k$-th DWF interval (\textup{$k=1,2,...,N^{\textrm{D}}$})
are \textup{
\[
\mathfrak{T}_{k}^{\textrm{S}}=\Bigl[t_{k}^{\textrm{D}},t_{k}^{\textrm{D}}+T{}_{k}^{\textrm{S}}\Bigr),\phantom{=}\mathfrak{T}_{k}^{\textrm{R}}=\Bigl[t_{k}^{\textrm{D}}+T{}_{k}^{\textrm{S}},t_{k+1}^{\textrm{D}}\Bigr),
\]
}respectively. The transmit powers in the $k$-th DWF interval (\textup{$k=1,2,...,N^{\textrm{D}}$})
are

\textup{
\[
P_{k}^{\textrm{S}}=\frac{E{}_{k}^{\textrm{D}}}{T{}_{k}^{\textrm{S}}},\phantom{=}P_{k}^{\textrm{R}}=\frac{E_{k}^{\textrm{R}}}{T{}_{k}^{\textrm{D}}-T_{k}^{\textrm{S}}},
\]
}where \textup{$E_{k}^{\textrm{R}}$} and\textup{ $T{}_{k}^{\textrm{S}}$}
are obtained by solving Eqns. (\ref{eq:13})\textasciitilde{}(\ref{eq:15})
in Appendix 2\textup{. }\end{cor}
\begin{IEEEproof}
The proof is given in Appendix 2.
\end{IEEEproof}

Next, we consider the case with a power constrained relay, for which
another property that is stronger than Property 1 in Theorem \ref{thm:Property}
can be found, given in the following Lemma. 
\begin{lem}
\label{lem:power}Assume that the source follows a DWF EH profile
and the relay is power constrained. Then, when the optimal transmission
policy is adopted for \textbf{Problem} \textbf{RMAX}, \textup{$P{}^{\textrm{R}}\left(t\right)$}
can only be $0$ or \textup{$P_{M}^{\textrm{R}}$}.\end{lem}
\begin{IEEEproof}
As the relay transmit power is limited by the peak power constraint,
at any time transmitting below the peak power will degrade the performance.
Hence the relay uses a constant transmit power, which is the peak
power. 
\end{IEEEproof}

According to Theorem \ref{thm:Property} and Lemma \ref{lem:power},
the optimal solution with a power constrained relay is provided in
the following Corollary.
\begin{cor}
\textup{\label{cor:Solution power}}In the optimal transmission policy
for \textbf{Problem} \textbf{RMAX} with the DWF EH profile and a power
constrained relay,\textup{ }there is a single S-R stage pair in each
DWF interval, i.e., the source and relay stages for the $k$-th DWF
interval (\textup{$k=1,2,...,N^{\textrm{D}}$}) are \textup{
\[
\mathfrak{T}_{k}^{\textrm{S}}=\Bigl[t_{k}^{\textrm{D}},t_{k}^{\textrm{D}}+T{}_{k}^{\textrm{S}}\Bigr),\phantom{=}\mathfrak{T}_{k}^{\textrm{R}}=\Bigl[t_{k}^{\textrm{D}}+T{}_{k}^{\textrm{S}},t_{k+1}^{\textrm{D}}\Bigr),
\]
}respectively; and the transmit powers in the $k$-th DWF interval
(\textup{$k=1,2,...,N^{\textrm{D}}$}) are

\textup{
\[
P_{k}^{\textrm{S}}=\frac{E_{k}^{\textrm{D}}}{T_{k}^{\textrm{S}}},\phantom{=}P_{k}^{\textrm{R}}=P_{M}^{\textrm{R}},
\]
}where \textup{$T{}_{k}^{\textrm{S}}$} is the solution of the following
equation

\textup{
\begin{equation}
T_{k}^{\textrm{S}}g\left(\frac{E_{k}^{\textrm{D}}}{T_{k}^{\textrm{S}}}\right)=\left(T_{k}^{\textrm{D}}-T_{k}^{\textrm{S}}\right)g\left(P_{M}^{\textrm{R}}\right).\label{eq:7}
\end{equation}
}\end{cor}
\begin{IEEEproof}
We can see that Eqn. (\ref{eq:7}) satisfies all the conditions in
Theorem \ref{thm:Property} and Lemma \ref{lem:power}, and has a
unique solution, so its solution is exactly the optimal source period.
The solution of the relay periods and source powers can be easily
verified.
\end{IEEEproof}

Corollaries \ref{cor:Solution energy} and \ref{cor:Solution power}
provide optimal solutions for the RMAX Problem with an energy constrained
relay and a power constrained relay, respectively. For the case where
the relay has both energy and power constraints\textit{}%
\footnote{Note that when the peak power constraint is small, e.g., for a sensor
node, the power constraint will be the only constraint, while the
total energy constraint will not be tight.%
}, the optimal solution can be obtained by Algorithm \ref{alg:two_constr_relay}.

\begin{algorithm}
1)\ \ $P_{M}^{\textrm{R}*}\leftarrow P_{M}^{\textrm{R}}$, $P_{M}^{\textrm{R}}\leftarrow\infty$.

\ \ \ \ Obtain Solution 1 for \textbf{Problem RMAX }by \textbf{ }Corollary
\ref{cor:Solution energy}.

\ \ \ \ \textbf{If} $P_{k}^{\textrm{R}}\leq P_{M}^{\textrm{R}}$
holds for all $k=1$: $N^{\textrm{D}}$ \{Only the energy constraint
is tight.\}

\ \ \ \ \ \ \ \ Solution 1 is the optimal solution.

\ \ \ \ \ \ \ \ \textbf{Return}.

\ \ \ \ \textbf{End if}

2)\ \ $E_{\Sigma}^{\textrm{R}*}\leftarrow E_{\Sigma}^{\textrm{R}}$,
$E_{\Sigma}^{\textrm{R}}\leftarrow\infty$, $P_{M}^{\textrm{R}}\leftarrow P_{M}^{\textrm{R*}}$.

\ \ \ \ Obtain solution 2 for \textbf{Problem RMAX }by \textbf{ }Corollary
\ref{cor:Solution power}.

\ \ \ \ \textbf{If} $\sum_{k=1}^{N^{\textrm{D}}}\left(T_{k}^{\textrm{D}}-T_{k}^{\textrm{S}}\right)P_{k}^{\textrm{R}}\leq E_{\Sigma}^{\textrm{R}}$
\{Only the power constraint is tight.\}

\ \ \ \ \ \ \ \ Solution 2 is the optimal solution.

\ \ \ \ \ \ \ \ \textbf{Return}.

\ \ \ \ \textbf{End if}

3)\ \ $E_{\Sigma}^{\textrm{R}}\leftarrow E_{\Sigma}^{\textrm{R}*}$.\ \ \ \ \{Both
constraints are tight.\}

\ \ \ \ \textbf{For} $k=1$: $N^{\textrm{D}}-1$

\ \ \ \ \ \ \ \ $P_{i}^{\textrm{R}}\leftarrow P_{M}^{\textrm{R}}$
for\textbf{ }$i=N^{\textrm{D}}+1-k$: $N^{\textrm{D}}$.

\ \ \ \ \ \ \ \ $E_{\Sigma}^{\textrm{R}}\leftarrow E_{\Sigma}^{\textrm{R}}-P_{M}^{\textrm{R}}\sum_{i=N^{\textrm{D}}+1-k}^{N^{\textrm{D}}}\left(T_{i}^{\textrm{D}}-T_{i}^{\textrm{S}}\right)$.

\ \ \ \ \ \ \ \ Obtain solution for \textbf{Problem RMAX }by
\textbf{ }Corollary \ref{cor:Solution energy} with $j=1$: $N^{\textrm{D}}-k$.

\ \ \ \ \ \ \ \ \textbf{If} $P_{j}^{\textrm{R}}\leq P_{M}^{\textrm{R}}$
holds for $j=1$: $N^{\textrm{D}}-k$

\ \ \ \ \ \ \ \ \ \ \ \ The current solution is the optimal
solution\textbf{.}

\ \ \ \ \ \ \ \ \ \ \ \ \textbf{Return.}

\ \ \ \ \ \ \ \ \textbf{End if}

\ \ \ \ \textbf{End for}

\caption{\label{alg:two_constr_relay}\hspace{-0.0405in}\textbf{:} \hspace{0.0405in}Optimal
solution for the case where the source assumes a DWF EH profile and
the relay has both energy and power constraints. }

\end{algorithm}

Steps 1 and 2 of Algorithm \ref{alg:two_constr_relay} can be proved
to be optimal directly via Corollaries \ref{cor:Solution energy}
and \ref{cor:Solution power}. For Step 3, in each \textbf{``for''}
loop, it can be verified that the relay power equals the peak power
for all the $i=N^{\textrm{D}}+1-k$: $N^{\textrm{D}}$, by checking
previous steps and the monotonic property (see Lemma 4 in \cite{EH_two_hop2})
of the relay transmit power. Furthermore, according to the two Corollaries,
we can observe that once the ``\textbf{for}'' loop terminates, the
current solution is optimal.

\subsection{Achievability of the Throughput Upper Bound with the Original EH
Profile}

\begin{figure}
\begin{centering}
\includegraphics[scale=0.47]{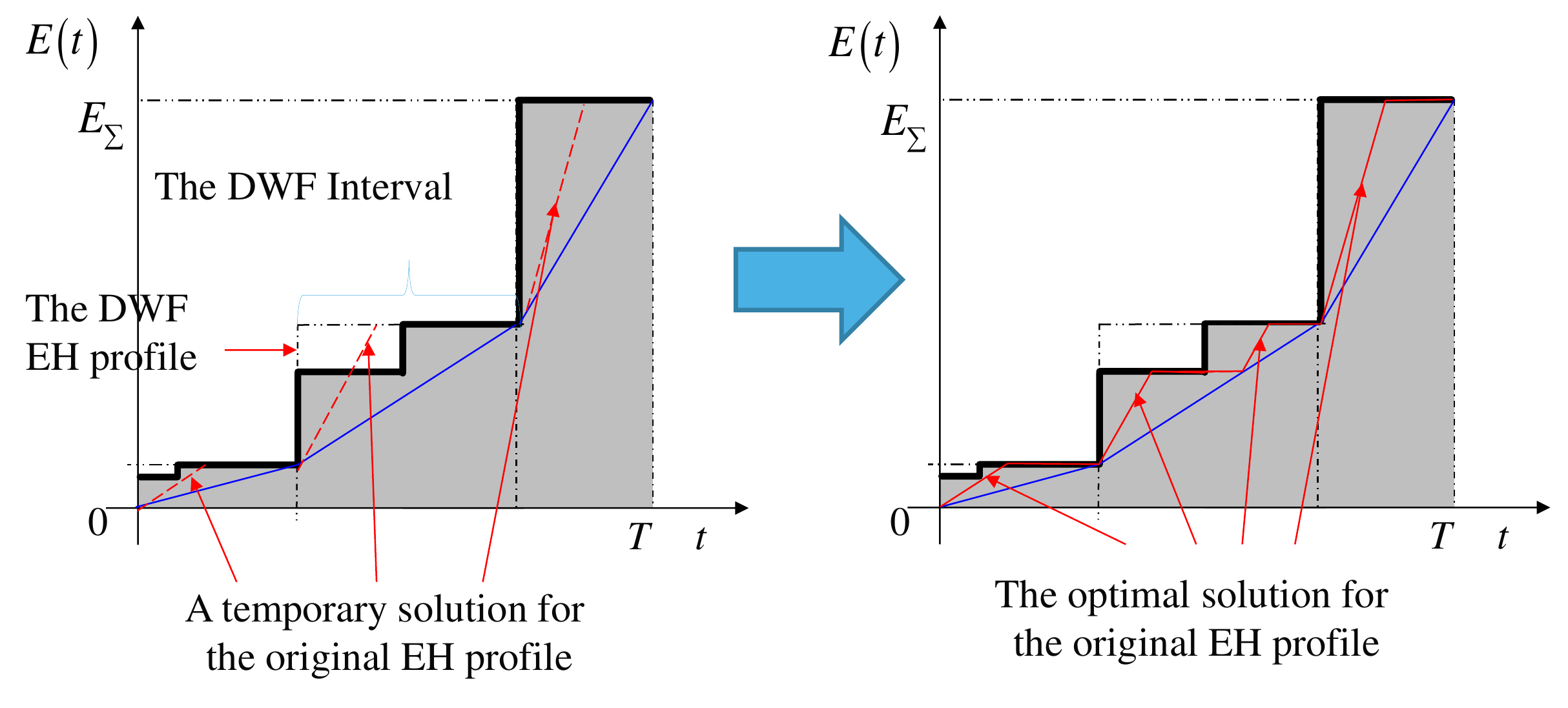}
\par\end{centering}

\caption{\label{fig:Obtaining-the-optimal}An illustration of how to modify
the solution for the DWF EH profile to get the optimal solution for
the original EH profile. A temporary solution and the optimal solution
are shown by the power allocation curves for the source, the slope
of each of which denotes the instantaneous source transmit power.
In this example, in the first and third DWF intervals, the temporary
solution is already the final solution. However, note that in the
second interval, the optimal solution is obtained through the iterative
steps described in the proof. }
\end{figure}

We have derived the optimal solution for the DWF EH profile, which
provides an upper bound of the achievable throughput for the original
EH profile. Next we will show how to achieve this upper bound. For
the original EH profile, the assumption that each DWF interval includes
one S-R stage pair may no longer hold, but by dividing each DWF interval
into more stages and carefully designing these stages, we can achieve
the throughput upper bound. Define the total source period for the
$k$-th DWF interval as the sum of all source periods in this DWF
interval (also the length of $\mathfrak{T}_{k}^{\textrm{S}}$), denoted
as $T_{k,\Sigma}^{\textrm{S}}$, then this main result is stated in
the following Theorem. 
\begin{thm}
\label{thm:achievablity}The maximum throughput of the considered
two-hop communication system with the original EH profile is the same
as that under its associated DWF EH profile. In addition, in each
DWF interval, the optimal solutions for \textbf{Problem RMAX} with
both EH profiles share the same values for the following parameters:
1) the source (relay) transmit power\textup{ $P_{k}^{\textrm{S}}$
}(\textup{$P_{k}^{\textrm{R}}$}), and 2) the total source period
\textup{$T_{k,\Sigma}^{\textrm{S}}$}.\end{thm}
\begin{IEEEproof}
It can be verified that for any two feasible solutions, as long as
they have the same parameters as stated in the Theorem, they achieve
the same throughput. Therefore, we need to find a feasible solution
for the original EH profile that has the same parameters as those
for the DWF EH profile. Without loss of generality, we consider the
$k$-th DWF interval, and treat the solution for the DWF EH profile
as a temporary solution. If the temporary power consumption curve
is inside the feasible domain of the original EH profile, then the
temporary solution is exactly the desired final solution. Otherwise,
the source will transmit with power $P_{k}^{\textrm{S}}$ until all
the available energy is used, and then the relay will transmit with
power $P_{k}^{\textrm{R}}$ until it empties its data buffer. We can
then restart the source transmission, and iterate as in the previous
steps until we reach the end of the DWF interval. This process is
illustrated in Fig. \ref{fig:Obtaining-the-optimal}. It can be easily
verified that the new scheduling enjoys the same set of critical parameters
as the ones for the DWF EH profile and it does not violate the data
or energy constraint.
\end{IEEEproof}

Based on Algorithm \ref{alg:two_constr_relay} and Theorem \ref{thm:achievablity},
the optimal solution for the RMAX problem with the original EH profile
can be obtained by Algorithm \ref{alg::solution_RMAX}. Note that
all the $i$, $j$, $k$ in the algorithm are integers. If the set
of $j$, i.e., $U_{\left\{ j\right\} }$, is an empty set, then no
$j$ violates the required condition. We omit these two statements
in the rest of this paper.

\begin{algorithm}
1) Adopt Algorithm \ref{alg:two_constr_relay} to obtain: DWF points
$\left(t_{k}^{\textrm{D}},\sum_{i=1}^{k}E{}_{i}^{\textrm{D}}\right)$;
solutions of $T{}_{k}^{\textrm{S}}$, $P{}_{k}^{\textrm{S}}$, $P{}_{k}^{\textrm{R}}$
; indices $i_{k}$ with $k=1$: $N^{\textrm{D}}$.

2)\textbf{ For} $k=1$: $N^{\textrm{D}}$

\ \ \ \ \ \ $E_{0}^{\tau}\leftarrow\sum_{l=1}^{i_{k-1}}E_{l}$,
$\tau_{0}\leftarrow t_{i_{k-1}}$, $m_{k}\leftarrow1$, $T_{k,1}^{\textrm{S}}\leftarrow T_{k}^{\textrm{S}}$, 

\ \ \ \ \ \ $U_{\left\{ j\right\} }\leftarrow\left\{ j\left|j\leq i_{k}-1,t_{j}-\tau_{0}>0\right.\right\} $.

\ \ \ \ \ \ \textbf{While} $I\left(j\right)=P_{k}^{\textrm{S}}\left(t_{j}-\tau_{0}\right)-\left(\sum_{l=1}^{j}E_{l}-E_{0}^{\tau}\right)\leq0$
does not hold for all the $j\in U_{\left\{ j\right\} }$

\ \ \ \ \ \ \{The solution for the DWF EH profile is not feasible
for the original EH profile.\}

\ \ \ \ \ \ \ \ \ \ $j'\leftarrow\min\{j\left|j\in U_{\left\{ j\right\} },I\left(j\right)>0\right.\}$,
$E_{0}^{\tau}\leftarrow\sum_{l=1}^{j'}E_{l}$ , $T_{k,m_{k}}^{\textrm{S}}\leftarrow\frac{E_{0}^{\tau}}{P_{k}^{\textrm{S}}}$.

\ \ \ \ \ \ \ \ \ \ \{Obtain actual solution by dividing
each DWF interval into multiple S-R stage pairs.\}

\ \ \ \ \ \ \ \ \ \ Update $U_{\left\{ j\right\} }$ with
the new $\tau_{0}\leftarrow T_{k,m_{k}}^{\textrm{S}}\frac{T_{k}^{\textrm{D}}}{T_{k}^{\textrm{S}}}$.

\ \ \ \ \ \ \ \ \ \ $m_{k}\leftarrow m_{k}+1$.

\ \ \ \ \ \ \textbf{End while}

\ \ \textbf{End for}

For the $k$-th DWF interval, the solutions of the source (or relay)
periods are $\left(T_{k,1}^{\textrm{S}},T_{k,2}^{\textrm{S}},...,T_{k,m_{k}}^{\textrm{S}}\right)$
(or $\frac{T_{k}^{\textrm{D}}-T_{k}^{\textrm{S}}}{T_{k}^{\textrm{S}}}\left(T_{k,1}^{\textrm{S}},T_{k,2}^{\textrm{S}},...,T_{k,m_{k}}^{\textrm{S}}\right)$),
while the source (or relay) power is $P_{k}^{\textrm{S}}$ (or $P_{k}^{\textrm{R}}$).

\textbf{Return}.

\caption{\label{alg::solution_RMAX}\hspace{-0.0405in}\textbf{:} \hspace{0.0405in}The
optimal solution for \textbf{Problem RMAX}}
\end{algorithm}

The procedures in Algorithm \ref{alg::solution_RMAX} can be divided
into two parts. The initial step is to obtain the solution for the
DWF EH profile, which can serve as a temporary solution for the original
EH profile. The objective of Step 2 is to obtain the actual solution
based on the temporary solution. Also note that we actually only need
to determine the scheduling, as the power allocation has been completed
in Step 1. 
\begin{rem}
We see that the DWF power allocation result for the single-hop transmission,
which determines the DWF EH profile, can serve as a design guideline
for the two-hop communication system. First, it determines the value
of the optimal throughput. Second, it allows us to decouple the joint
scheduling and power allocation problem, and provides the optimal
power allocation solution eventually. This indicates that the DWF
power allocation plays a central role in EH communication systems.
\end{rem}

\begin{rem}
Though we only consider the staircase-shaped EH profiles in this paper,
this algorithm can be directly applied for more general EH profiles.
In fact, for an EH source node, whether the EH profile is continuous
or discrete, as long as its cumulative EH curve is between the DWF
EH profile and the DWF power allocation curve, its optimal transmission
policy is provided by Algorithm \ref{alg::solution_RMAX} and its
throughput is the same with that of the DWF EH profile. 
\end{rem}

\section{Transmission Completion Time Minimization }

Problem TMIN is the dual problem of RMAX, where the total transmission
time is unknown and needs to be minimized. For the single-hop case,
an algorithm%
\footnote{In this section, the DWF algorithm refers to this particular algorithm
(i.e., not the one we used in Section III).%
} similar to DWF was proposed in \cite{EH_DWF_time} to minimize the
transmission time. Unfortunately, in contrast to RMAX, the result
of the single-hop DWF power allocation for TMIN cannot provide the
respective DWF EH profile that we need in the two-hop case. Hence,
the main challenge is to obtain the required DWF points, and then
the algorithm for RMAX can be applied for the scheduling and power
allocation.

\subsection{Obtaining All the DWF Points with a Power Constrained Relay}

We will first consider a power constrained relay. From Theorem \ref{thm:achievablity},
we can see that with the optimal scheduling and power allocation,
the same amount of data can be transmitted with either the DWF EH
profile or the original EH profile, i.e., only the DWF points will
affect the transmission completion time. According to Theorems \ref{lem:power}
and \ref{thm:achievablity}, along with the properties of the DWF
power allocation \cite{EH_DWF_time}, we are now ready to characterize
the DWF points for the TMIN Problem.
\begin{lem}
\label{lem:DWF points}For a given amount of data $D$ and its associated
minimum transmission completion time $T$, the following equation
set is the necessary and sufficient condition for the set of points
\textup{$\left(t_{i_{k}},\sum_{j=1}^{i_{k}}E_{j}\right)$} (\textup{$k$=1,2,...,$N^{\textrm{D}}$})
to be the DWF points of the given EH profile:\textup{
\begin{equation}
\sum_{k=1}^{N^{\textrm{D}}}T_{k}^{\textrm{S}}g\left(P_{k}^{\textrm{S}}\right)=D,\label{eq:8}
\end{equation}
}and for $k$=1,2,...,\textup{$N^{\textrm{D}}-1$,
\begin{gather}
i_{k}=\arg\underset{i:s_{i}\leq T}{\min}\left\{ \frac{\sum_{j=i_{k-1}+1}^{i}E_{j}}{t{}_{i}-t_{i_{k-1}}}\right\} ,\label{eq:9}\\
T_{k}^{\textrm{D}}=t_{i_{k}}-t_{i_{k-1}},\begin{array}{c}
\end{array}P_{k}^{\textrm{S}}=\frac{\sum_{j=i_{k-1}+1}^{i_{k}}E_{j}}{t_{i_{k}}-t_{i_{k-1}}},\label{eq:10}\\
T_{k}^{\textrm{S}}g\left(P_{k}^{\textrm{S}}\right)=\left(T_{k}^{\textrm{D}}-T_{k}^{\textrm{S}}\right)g\left(P_{M}^{\textrm{R}}\right).\label{eq:11}
\end{gather}
}\end{lem}
\begin{IEEEproof}
(\ref{eq:9}) is due to the DWF algorithm (See (\ref{eq:6})), while
(\ref{eq:8}) is due to the transmission data constraint. Finally,
(\ref{eq:10}) and (\ref{eq:11}) can be obtained directly from Corollary
\ref{cor:Solution power}.
\end{IEEEproof}
Lemma \ref{lem:DWF points} provides the necessary and sufficient
condition for a set of points to be the DWF points. However, the optimization
in condition (\ref{eq:9}) requires the value of $T$, which can not
be determined before the optimization is finished, i.e., the condition
(\ref{eq:9}) needs a non-causal knowledge, which makes it difficult
to obtain the DWF points through Lemma \ref{lem:DWF points}. 

Next we will show that this non-causal requirement can be removed.
According to Properties 1 and 3 of DWF, the first DWF point is the
point where the source transmit power changes for the first time.
To obtain the first DWF point, we only need to replace $T$ in (\ref{eq:9})
with an earlier deadline before which the source transmit power must
have changed at least once. All the other DWF points can be obtained
iteratively using a similar approach by updating the remaining energy
and data. Based on this, we propose Algorithm \ref{alg:TMIN_power}
to find all the DWF points and also the transmission completion time.

\begin{algorithm}
$i_{0}\leftarrow0$, $k\leftarrow1$, $T^{\textrm{R}}\leftarrow\frac{D}{g\left(P_{M}^{\textrm{R}}\right)}$,
$j_{0}\leftarrow\min\left\{ j\left|s_{j}>T^{\textrm{R}}\right.\right\} $.

\textbf{While }$D>0$

\ \ \ \ $E_{k,j}^{\textrm{req}}\leftarrow\left(t_{j}-T^{\textrm{R}}\right)g^{-1}\left(\frac{D}{t_{j}-T^{\textrm{R}}}\right)$,
with $j=j_{0}$: $N$

\ \ \ \ Obtain $\tilde{i_{k}}$ and $t_{k}'$ by $\tilde{i_{k}}=\min\left\{ j\left|E_{k,j}^{\textrm{req}}\leq\sum_{l=1}^{j}E_{l}\right.\right\} $
and $g\left(\frac{\sum_{i=1}^{\tilde{i_{k}}}E_{i}}{t_{k}'-T^{\textrm{R}}}\right)=\frac{D}{t_{k}'-T^{\textrm{R}}}$.

\ \ \ \ \{$\tilde{i_{k}}$ is the previously-mentioned deadline
for the $k$-th DWF point.\}

\ \ \ \ $T_{k}^{\textrm{D}}\leftarrow t_{k}',$ $E_{k}^{\textrm{D}}\leftarrow\sum_{i=1}^{\tilde{i_{k}}}E_{i}$,
$T_{k}^{\textrm{R}}\leftarrow T^{\textrm{R}}$.

\begin{raggedright}
\ \ \ \ \textbf{If }$\frac{\sum_{i=1}^{\tilde{i_{k}}}E_{i}}{t_{k}'}t_{j}\leq\sum_{l=1}^{j}E_{l}$
, $j\in U_{\left\{ j\right\} }=\left\{ j\left|0<j<\tilde{i_{k}}\right.\right\} $
\par\end{raggedright}

\begin{raggedright}
\ \ \ \ \{Constant power transmission is permitted by the DWF
EH profile before $\tilde{i_{k}}$.\}
\par\end{raggedright}

\begin{raggedright}
\ \ \ \ \ \ \ \ \textbf{Break.}
\par\end{raggedright}

\begin{raggedright}
\ \ \ \ \textbf{End if}
\par\end{raggedright}

\ \ \ \ $i_{k}\leftarrow\arg\underset{i<\tilde{i_{k}}}{\min}\left\{ \frac{\sum_{j=i_{k-1}+1}^{i}E_{j}}{t{}_{i}-t_{i_{k-1}}}\right\} ,$
$T_{k}^{\textrm{D}}\leftarrow t_{i_{k}}$, $E_{k}^{\textrm{D}}\leftarrow\sum_{j=i_{k-1}+1}^{i_{k}}E_{j}$.

\ \ \ \ \{$i_{k}$ is the $k$-th DWF point.\}

\ \ \ \ $T_{k}^{\textrm{R}}\leftarrow$ $\left(T_{k}^{\textrm{D}}-T_{k}^{\textrm{R}}\right)g\left(\frac{E_{k}^{\textrm{D}}}{T_{k}^{\textrm{D}}-T_{k}^{\textrm{R}}}\right)=T_{k}^{\textrm{R}}g\left(P^{\textrm{R}}\right)$.

\ \ \ \ $t_{j-i_{k}}\leftarrow t_{j}-t_{i_{k}}$, $E_{j-i_{k}}\leftarrow E_{j}$,
for all the $i_{k}\leq j\leq N$.

\ \ \ \ $D\leftarrow D\frac{T^{\textrm{R}}-T_{k}^{\textrm{R}}}{T^{\textrm{R}}}$,
$T^{\textrm{R}}\leftarrow T^{\textrm{R}}-T_{k}^{\textrm{R}}$, $k\leftarrow k+1$.

\ \ \ \ \{All parameters are updated for obtaining the remaining
DWF points.\}

\textbf{End while}

$N^{\textrm{D}}=k$, the transmission completion time is $T=t_{N^{\textrm{D}}}^{\textrm{D}}$,
coordinates of DWF points are $\left(\sum_{j=1}^{k}T_{j}^{\textrm{D}},\sum_{j=1}^{k}E_{j}^{\textrm{D}}\right)$.

\textbf{Return}.

\caption{\label{alg:TMIN_power}\hspace{-0.0405in}\textbf{:} \hspace{0.0405in}Finding
all the DWF points and the transmission completion time for the power
constrained relay case. }

\end{algorithm}

\begin{thm}
\label{thm:All-the-DWF}All the DWF points found in Algorithm \ref{alg:TMIN_power}
satisfy the conditions in Lemma \ref{lem:DWF points}.\end{thm}
\begin{IEEEproof}
The proof for Theorem \ref{thm:All-the-DWF} is given in Appendix
3.\end{IEEEproof}
\begin{rem}
Algorithm \ref{alg:TMIN_power} not only obtains all the DWF points
and the transmission completion time $T=t_{N^{\textrm{D}}}^{\textrm{D}}$,
but also determines the optimal power allocation ($P_{k}^{\textrm{S}}=\frac{E_{k}^{\textrm{D}}}{T_{k}^{\textrm{D}}-T_{k}^{\textrm{R}}}$)
and scheduling ($T_{k}^{\textrm{S}}$ and $T_{k}^{\textrm{R}}$) for
the DWF EH profile, i.e., the temporary solution for the original
EH profile. 
\end{rem}

\subsection{Obtaining All the DWF Points with an Energy Constrained Relay}

For the energy constrained relay case, the relay cannot always transmit
with a constant power, which adds to the difficulty of optimization.
Therefore, Algorithm \ref{alg:TMIN_power} cannot be directly applied.
However, we can first force the relay to transmit with a constant
power in the whole transmission period to obtain an upper bound for
the transmission time. We can then find the optimal transmission time
between this upper bound and the beginning time. The algorithm of
obtaining all the DWF points for the energy constrained relay case
is given in Algorithm \ref{alg:TMIN_energy}, in which $R\left(k\right)$
denotes the throughput by adopting Corollary \ref{cor:Solution energy}
with $T\leftarrow\sum_{i=1}^{k}T_{i}^{\textrm{D}}$, while $R'\left(t\right)$
denotes the throughput by adopting Corollary \ref{cor:Solution energy}
and the single-hop DWF in (\ref{eq:9}) with $T\leftarrow t$.

\begin{algorithm}
Obtain $P^{\textrm{R}}$ by $\frac{E_{\Sigma}^{\textrm{R}}}{P^{\textrm{R}}}g\left(P^{\textrm{R}}\right)=D$.

Obtain all the $N^{\textrm{D}'}$ temporary DWF points by Algorithm
\ref{alg:TMIN_power}.

$k_{l}\leftarrow0$, $k_{u}\leftarrow N^{\textrm{D}'}$.

Find $\hat{k}$ that satisfies $R\left(\hat{k}\right)\leq D$ and
$R\left(\hat{k}+1\right)>D$ using bisection.

\{The $\hat{k}$-th DWF interval contains the ending time.\}

\textbf{If} $R\left(\hat{k}\right)=D$

\ \ \ \  $T\leftarrow\sum_{i=1}^{\hat{k}}T_{i}^{\textrm{D}}$.

\textbf{Else}

\ \ \ \ $t_{l}\leftarrow\sum_{i=1}^{\hat{k}}T_{i}^{\textrm{D}}$,
$t_{u}\leftarrow\sum_{i=1}^{\hat{k}+1}T_{i}^{\textrm{D}}$.

\ \ \ \ For $t\in\left[t_{l},t_{u}\right)$, use bisection to
find $\hat{t}$ that makes $R'\left(\hat{t}\right)=D$.

\ \ \ \ $T\leftarrow\hat{t}$. \{This step is to decide the exact
value of the ending time.\}

\textbf{End}

$T$ is the transmission completion time, the DWF points and solutions
are obtained by Corollary \ref{cor:Solution energy} for this $T$.

\textbf{Return}.

\caption{\label{alg:TMIN_energy}\hspace{-0.0405in}\textbf{:} \hspace{0.0405in}Finding
all the DWF points and the transmission completion time for the energy
constrained relay case. }

\end{algorithm}

\begin{rem}
A similar procedure as Algorithm \ref{alg:TMIN_energy} can be derived
to obtain the DWF points for the case where the relay has both power
and energy constraints.
\end{rem}

\subsection{Optimal Transmission Policy}

With Algorithms \ref{alg:TMIN_power} and \ref{alg:TMIN_energy},
the optimal solution to TMIN with the original EH profile can be found
as follows:

1) Obtain all the DWF points $\left(t_{k}^{\textrm{D}},\sum_{i=1}^{k}E_{i}^{\textrm{D}}\right)$,
as well as the respective parameters of $P_{k}^{\textrm{S}}$ ($P_{k}^{\textrm{R}}$),
$T_{k}^{\textrm{S}}$ ($T_{k}^{\textrm{R}}$) for $1\leq k\leq N^{\textrm{D}}$,
by Algorithm \ref{alg:TMIN_energy}.

2) Obtain the optimal transmission policy by Step 2 of Algorithm \ref{alg::solution_RMAX}. 

Similar to the RMAX problem, the DWF EH profile serves as the building
block for solving the TMIN problem. This novel approach to study the
two-hop EH communication system, by first investigating the DWF EH
profile and then extending the result to the original EH profile,
has the potential to solve other design problems.

\section{Simulation Results}

In this section, we provide simulation results to demonstrate the
performance of our proposed algorithms and to show the importance
of both scheduling and power allocation optimizations in two-hop EH
communication systems.

\subsection{Short-term Throughput Maximization }

We first investigate the throughput maximization problem. To verify
the importance of both the scheduling and power allocation, we consider
two baseline policies: Fixed scheduling and fixed power allocation.
For the first policy, we fix a two-equal-stage scheduling but adopt
the optimal power allocation in each stage. In the second one, we
adopt the optimal scheduling locally inside each energy arrival interval
so that the source and the relay transmit the same amount of data,
but a fixed transmit power allocation is applied at the source. All
the policies are also compared with a throughput upper bound, assuming
a non-EH source with the same amount of total energy. 

\begin{figure}
\begin{centering}
\includegraphics[scale=0.72]{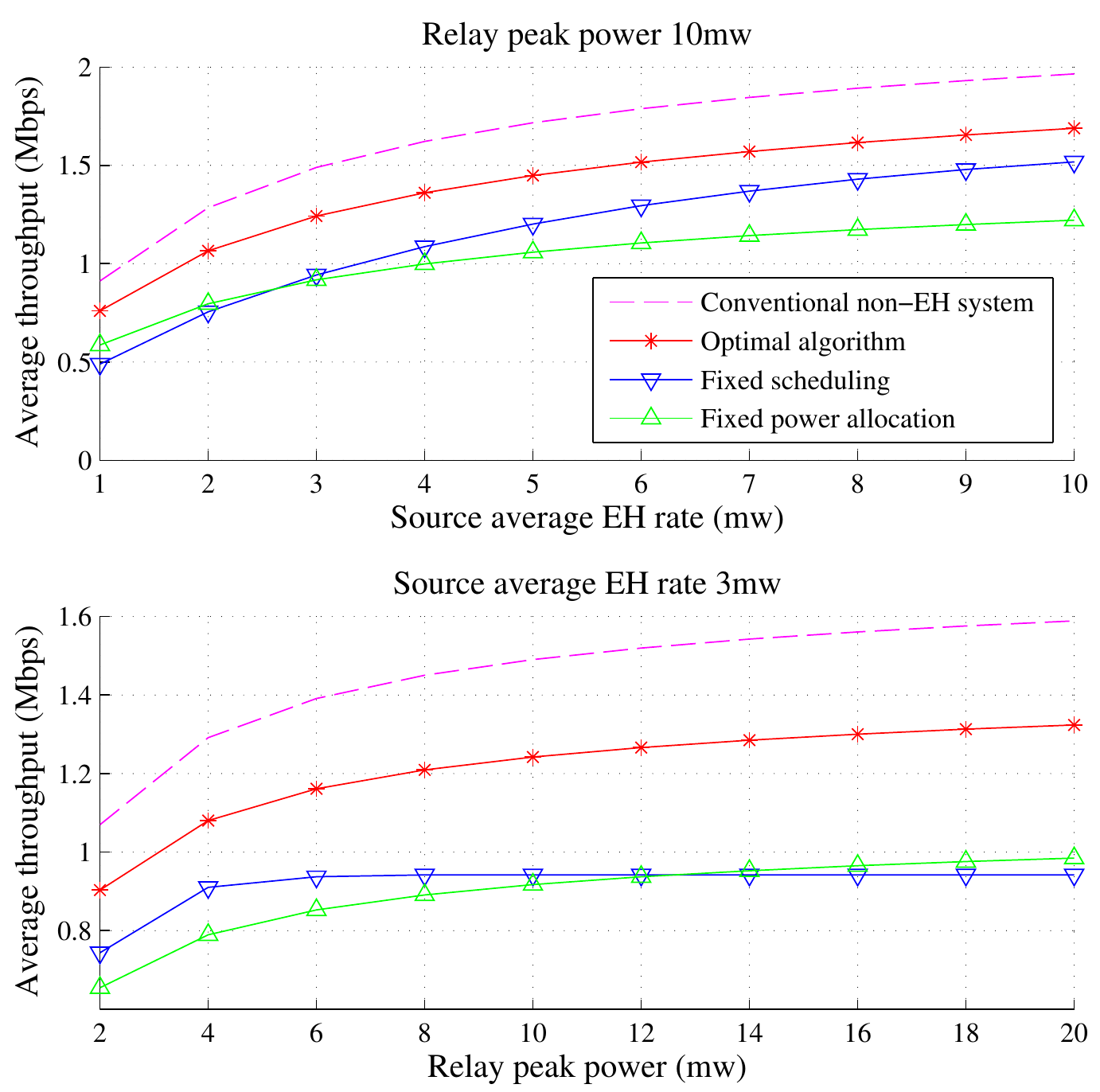}
\par\end{centering}

\caption{\label{fig:Simulation-results-of}Simulation results of the throughput
versus the source average EH rate and the relay peak power respectively.}
\end{figure}

We consider a band-limited additive white Gaussian noise channel,
with bandwidth $W$ = 1MHz and noise power spectral density $N_{0}=10^{\text{\textminus}19}$W/Hz.
We assume that the path loss $H$ is 100dB, and the transmission block
length is 100ms. The rate-power function is $r=W\log_{2}\left(1+\frac{PH}{N_{0}W}\right)=\log_{2}\left(1+\frac{P}{10^{-3}}\right)\textrm{Mbps}$.
The number of energy arrivals in a time interval of length $t$ follows
a Poisson distribution with mean $\lambda_{e}t$ and is in the unit
of $E_{0}^{\textrm{EH}}$. The \emph{average EH rate} is then defined
as $P^{\textrm{EH}}=E_{0}^{\textrm{EH}}\lambda_{e}$, e.g., if $E_{0}^{\textrm{EH}}=1\textrm{mJ}$
and $\lambda_{e}=1\textrm{sec}^{-1}$, then $P^{\textrm{EH}}=1\textrm{mJ}/\textrm{sec}=1\textrm{mw}$.
The simulation is run for 1000 random EH realizations with $\lambda_{e}=1\textrm{sec}^{-1}$.
Both the throughput versus the source average EH rate with the relay
peak power $P_{M}^{\textrm{R}}=10\textrm{mw}$ and the throughput
versus the relay peak power with the average EH rate $P^{\textrm{EH}}=3\textrm{mw}$
are shown in Fig. \ref{fig:Simulation-results-of}. The numerical
values of the different simulation parameters are selected based on
the EH wireless sensor STM 11x \cite{website:enocean}, which can
harvest energy at the rate of $3\textrm{mw}$ with a peak transmit
power of 10mw.

We see that there is a large throughput gap between our optimal solution
and the two baseline policies, and there is a performance loss compared
to the upper bound due to the random EH profile. The first baseline
case is constrained by the fixed scheduling, as the equal-length stages
cannot adapt to different EH levels of the source. Therefore, even
when the source has enough energy and can spend less time to transmit,
the source stage will still occupy a fixed period. The second baseline
policy also performs poorly, and this is because the power allocation
is not adapted to the source EH profile. These two observations demonstrate
the importance of both the optimal scheduling and power allocation
for two-hop EH systems. We shall also mention that the performance
of both fixed policies is always not good as long as the source EH
profile varies with time. Meanwhile, the fixed scheduling policy performs
especially poorly when the source and relay power levels differ significantly.

\subsection{Transmission Completion Time Minimization }

For the transmission completion time minimization, we will also compare
the optimal policy with two baseline policies along with a performance
upper bound. All the other settings are the same as those of the RMAX
Problem except that in the first baseline policy, the only two stages
are not of the same length, but in each of them the same amount of
data is transmitted.

\begin{figure}
\begin{centering}
\includegraphics[scale=0.72]{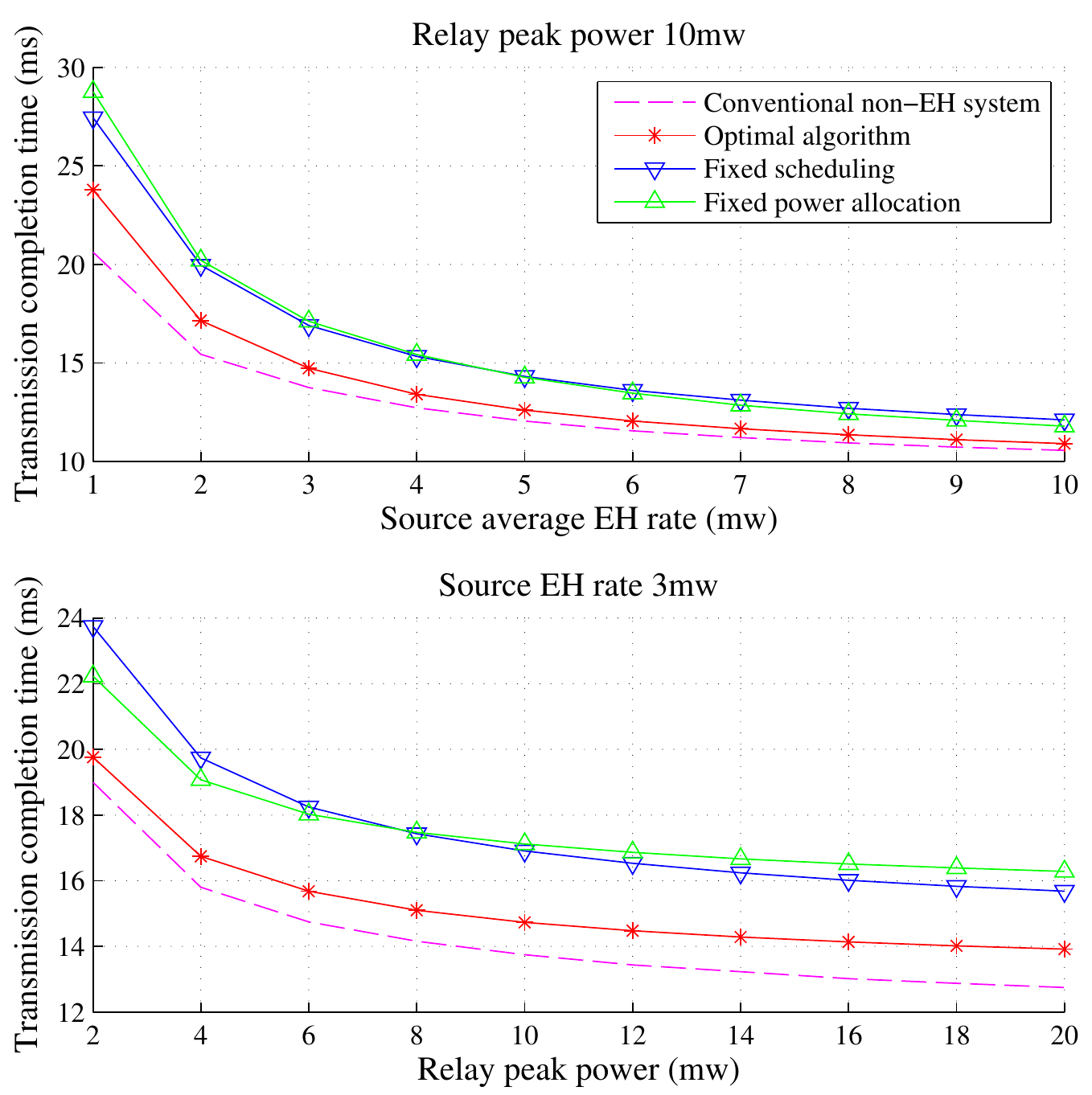}
\par\end{centering}

\caption{\label{fig:Simulation-results-of-1}Simulation results of the transmission
completion time versus the source average EH rate and the relay peak
power respectively.}
\end{figure}

The simulation is run for 1000 random EH realizations, with the same
parameter setting as that of Fig. \ref{fig:Simulation-results-of}.
The total amount of data that needs to be transmitted is 20kbits.
The transmission completion time versus the source average EH rate
with the relay peak power $P_{M}^{\textrm{R}}=10\textrm{mw}$ and
the transmission completion time versus the relay peak power with
the average EH rate $P^{\textrm{EH}}=3\textrm{mw}$ are shown in Fig.
\ref{fig:Simulation-results-of-1}. Compared with the two baseline
policies, we can notice a great performance improvement when employing
the optimal scheduling and power allocation. Again, this result demonstrates
the importance of both of the optimal scheduling and power allocation.

\section{Conclusions}

In this paper, we investigated the optimal transmission policy for
two-hop communication systems with an EH source assisted by a non-EH
relay, by either maximizing the throughput with a fixed transmission
period or minimizing the transmission completion time with a fixed
amount of data. Compared to the case with an EH relay, the system
design, especially the S-R transmission scheduling, becomes more challenging
with an EH source. Our results have shown that the DWF power allocation
algorithm for single-hop EH communications provides useful design
insights in solving the problem. In particular, the proposed DWF EH
profile is a promising design tool for two-hop EH communication systems.

\section*{Appendix 1 \ \ \ \ \ \ \ Proof of Theorem \ref{thm:Property}}

We only provide the proof for the case where the relay is with a total
energy constraint, while the case where the relay has both power and
energy constraints can be similarly obtained.

\noindent \textbf{Property 1.} This property can be proved according
to Lemma 1 in \cite{EH_two_hop2}. Furthermore, we can conclude that
a transmission policy cannot be optimal if either the source or the
relay uses different transmit powers in any given DWF interval.

\vspace{10pt}

\noindent \textbf{Property 2.} For brevity and without loss of generality,
we only provide the proof for the first DWF interval with $N^{\textrm{D}}>1$,
while the proof for the other intervals can be similarly obtained.
The whole proof consists of the following three cases:

\noindent \emph{Case 1: The source transmit power changes in the second
DWF interval.} 

\begin{figure}
\begin{centering}
\includegraphics[scale=0.5]{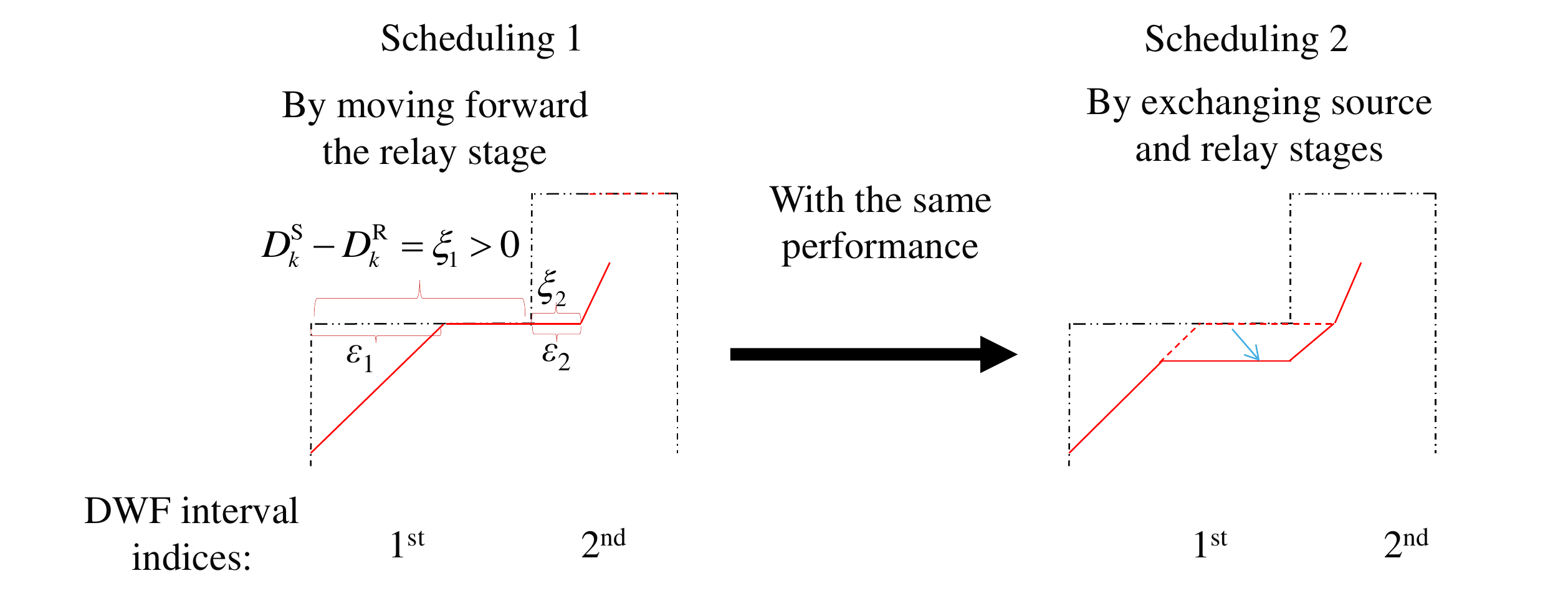}
\par\end{centering}

\caption{\label{fig:Illustration-of-the}Illustration of the proof of Case
1 for property 2 in Theorem \ref{thm:Property}. The solid curve represents
the power consumption at the source, i.e., the scheduling of the source,
and the dashed curve represents the power consumption curve of the
previous step. $\xi_{1}/\xi_{2}$ represent the amount of data, while
$\varepsilon_{1}/\varepsilon_{2}$ represent the time duration. }
\end{figure}

\noindent We define an \emph{equivalent scheduling} as the one that
achieves the same performance with the given scheduling decision but
without violating the feasibility condition. We will prove the relay
data equality by contradiction. Denote the last source period in the
first DWF interval as $\varepsilon_{1}$, and the amount of data transmitted
in the first relay stage after the first DWF interval as $\xi_{1}$.
Assume that the data equality does not hold and the amount of data
transmitted by the source is larger than the relay by an amount of
$\xi_{2}$ at the end of the first DWF interval. Then, we can always
find an equivalent scheduling, as Scheduling 1 in Fig. \ref{fig:Illustration-of-the},
where the beginning part of the second DWF interval is for the relay
stage, and the amount of data transmitted in it is $\xi=\textrm{min\ensuremath{\left(\xi_{1},\xi_{2}\right)}}$.
We denote this time period as $\varepsilon_{2}$. Next, from Scheduling
1, we can obtain another equivalent scheduling as Scheduling 2 in
Fig. \ref{fig:Illustration-of-the}, by exchanging the last time period
of $\varepsilon=\textrm{min\ensuremath{\left(\varepsilon_{1},\varepsilon_{2}\right)}}$
of the last source stage in the first DWF interval, with the relay
stage at the beginning of the second DWF interval. Note that previous
steps preserve the transmit powers for both the source and the relay.
After such re-scheduling, we can see that in the second DWF interval
of Scheduling 2, the source adopts a non-constant power, so we can
always find a new scheduling with a constant transmit power that achieves
a higher throughput, based on Property 1. Hence, the assumption does
not hold, and the data equality is proved.

\begin{figure}
\begin{centering}
\includegraphics[scale=0.5]{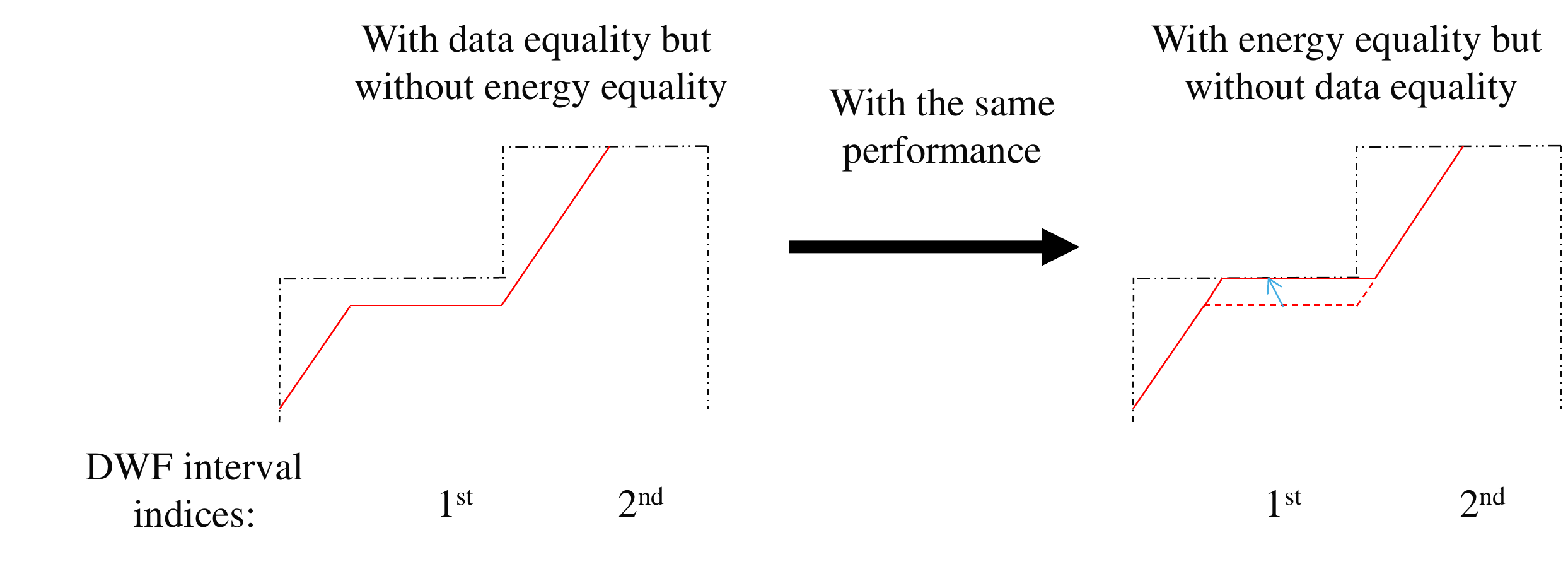}
\par\end{centering}

\caption{\label{fig:The-method-of}The method of obtaining the scheduling result
with the same performance but violating the data equality property,
for the proof of the source energy equality in Property 2 of Theorem
\ref{thm:Property}. }
\end{figure}

We will now prove the source energy equality with the help of the
relay data equality property. This part of the proof can be extended
to the other two cases. We assume that the source energy equality
does not hold. We can then always move the beginning part of the source
stage in the second DWF interval to the end of the source stage of
the first DWF interval. Meanwhile we can delay the relay stage so
that the source exhausts all the energy at the end of the first DWF
interval, as shown in Fig. \ref{fig:The-method-of}. This can be achieved
without violating either the energy or the data causality constraint,
while preserving the same throughput. It is obvious that the new scheduling
result violates the data equality property, and can always be replaced
by another policy with a better throughput, according to the previous
proof. The source energy equality is therefore also proved.

\noindent \emph{Case 2: In the second DWF interval, the relay transmit
power changes.}

\noindent According to Lemma 7 in \cite{EH_two_hop2}, when the relay
transmit power changes, at least one of the following situations happens:
The relay exhausts all the energy at that time instant, or the relay
has already transmitted all the data from the source. Otherwise, it
can always be replaced by another scheduling decision with a better
throughput. But for a non-EH relay it is obvious that the energy buffer
can only be empty at the end of the whole transmission period. Hence,
the relay must empty its data buffer. As such, the relay data equality
is proved.

\noindent \emph{Case 3: In the second DWF interval, both the source
and relay transmit powers do not change.} 

\noindent We will prove that this case in fact cannot exist. We denote
the index of the DWF interval at the end of which either the source
or relay transmit power changes for the first time as $k$. Denote
the time average of the source transmit power in $\left[0,t\right]$
as $P_{ave}\left(t\right)=\frac{\int_{0}^{t}P^{\textrm{S}}\left(\tau\right)d\tau}{t}$
for $t\in\left(0,t_{k}^{\textrm{D}}\right]$, then we can verify that
$P_{ave}\left(t_{1}^{\textrm{D}}\right)\geq\frac{E_{k}^{\textrm{S}}}{t_{k}^{\textrm{D}}}$.
However, due to Properties 1 and 3 of DWF, the total harvested energy
$E_{\Sigma}^{\textrm{EH}}\left(t_{1}^{\textrm{D}}\right)$ should
satisfy $\frac{E_{\Sigma}^{\textrm{EH}}\left(t_{1}^{\textrm{D}}\right)}{t_{1}^{\textrm{D}}}<\frac{E_{k}^{\textrm{S}}}{t_{k}^{\textrm{D}}}$
at the first DWF point. By combining these two aspects, we can obtain
$P_{ave}\left(t_{1}^{\textrm{D}}\right)>\frac{E_{\Sigma}^{\textrm{EH}}\left(t_{1}^{\textrm{D}}\right)}{t_{1}^{D}}$,
which contradicts the energy causality constraint that $E_{\Sigma}^{\textrm{EH}}\left(t_{1}^{\textrm{D}}\right)\geq E_{1}^{\textrm{S}}=P_{ave}\left(t_{1}^{\textrm{D}}\right)t_{1}^{\textrm{D}}$.
Hence, Case 3 is not feasible.

\section*{Appendix 2 \ \ \ \ \ \ \ Proof of Corollary \ref{cor:Solution energy}}

\noindent Based on Theorem \ref{thm:Property}, \textbf{Problem RMAX}
can be simplified with only equality constraints: 
\begin{align*}
\underset{{T_{k}^{\textrm{S}},E_{k}^{\textrm{R}}\atop k=1,2,...,N^{\textrm{D}}}}{\mathbf{\max}} & D{}^{\textrm{R}}\left(T\right)=\sum_{k=1}^{N^{\textrm{D}}}\left(T_{k}^{\textrm{D}}-T_{k}^{\textrm{S}}\right)g\left(\frac{E_{k}^{\textrm{R}}}{T_{k}^{\textrm{D}}-T_{k}^{\textrm{S}}}\right)\\
\textrm{\ensuremath{\phantom{==}}s.t.\ensuremath{\phantom{==}}} & T_{k}^{\textrm{S}}g\left(\frac{E_{k}^{\textrm{D}}}{T_{k}^{\textrm{S}}}\right)=\left(T_{k}^{\textrm{D}}-T_{k}^{\textrm{S}}\right)g\left(\frac{E_{k}^{\textrm{R}}}{T_{k}^{\textrm{D}}-T_{k}^{\textrm{S}}}\right),\\
 & \sum_{k=1}^{N^{\textrm{D}}}E_{k}^{\textrm{R}}=E_{\Sigma}^{\textrm{R}},
\end{align*}
In this new expression, there are $N^{\textrm{D}}$ variables $T_{k}^{\textrm{S}}$,
and $N^{\textrm{D}}$ variables $E_{k}^{\textrm{R}}$, $k=1,...,N^{\textrm{D}}$.
The equation of the first constraint for each $k$ in fact only includes
two variables: $T_{k}^{\textrm{S}}$ and $E_{k}^{\textrm{R}}$. We
can obtain a unique $T_{k}^{\textrm{S}}$ as a function of $E_{k}^{\textrm{R}}$
due to the monotonicity of the function $\ensuremath{g\left(\cdot\right)}$,
denoted as $T_{k}^{\textrm{S}}=f_{k}\left(E_{k}^{\textrm{R}}\right)$.
Although an explicit expression of the function $f_{k}$ is difficult
to obtain, we can check that it is a monotonically increasing function
by obtaining its inverse function $E_{k}^{\textrm{R}}=f_{k}^{-1}\left(T_{k}^{\textrm{S}}\right)=\left(T_{k}^{\textrm{D}}-T_{k}^{\textrm{S}}\right)g^{-1}\left(\frac{T_{k}^{\textrm{S}}}{T_{k}^{\textrm{D}}-T_{k}^{\textrm{S}}}g\left(\frac{E_{k}^{\textrm{D}}}{T_{k}^{\textrm{S}}}\right)\right)$.
Problem RMAX can then be simplified as an unconstrained problem with
the new objective function as 
\begin{align}
D{}^{\textrm{R}}\left(T\right) & =\overset{N^{\textrm{D}}-1}{\underset{k=1}{\sum}}\left(T_{k}^{\textrm{D}}-f_{k}\left(E_{k}^{\textrm{R}}\right)\right)g\left(\frac{E_{k}^{\textrm{R}}}{T_{k}^{\textrm{D}}-f_{k}\left(E_{k}^{\textrm{R}}\right)}\right)\label{eq:12}\\
 & +\left(T_{N^{\textrm{D}}}^{\textrm{D}}-f_{N^{\textrm{D}}}\left(E_{\Sigma}^{\textrm{R}}-\overset{N^{\textrm{D}}-2}{\underset{k=1}{\sum}}E_{k}^{\textrm{R}}\right)\right)g\left(\frac{E_{\Sigma}^{\textrm{R}}-\overset{N^{\textrm{D}}-2}{\underset{k=1}{\sum}}E_{k}^{\textrm{R}}}{T_{k}^{\textrm{D}}-f_{N^{\textrm{D}}}\left(E_{\Sigma}^{\textrm{R}}-\overset{N^{\textrm{D}}-2}{\underset{k=1}{\sum}}E_{k}^{\textrm{R}}\right)}\right)\nonumber 
\end{align}
 over only $\left(N^{\textrm{D}}-1\right)$ variables $E_{k}^{\textrm{R}}$.
It can be verified that $\frac{d^{2}D{}^{\textrm{R}}\left(T\right)}{d^{2}E_{k}^{\textrm{R}}}<0$,
for each given index $k$ and a given set of $\left\{ E_{i}^{\textrm{R}},\begin{array}{c}
\end{array}i=1,2,...,k-1,k+1,...,N^{\textrm{D}}-1\right\} $. Also $\frac{dD{}^{\textrm{R}}\left(T\right)}{dE_{k}^{\textrm{R}}}>0$
for $E_{k}^{\textrm{R}}=0$, and $\frac{dD{}^{\textrm{R}}\left(T\right)}{dE_{k}^{\textrm{R}}}<0$
for $E_{k}^{\textrm{R}}=E_{\Sigma}^{\textrm{R}}-\sum_{i=1,i\neq k}^{N^{\textrm{D}}-1}E_{i}^{\textrm{R}}$.
So the value of $E_{k}^{\textrm{R}}$ that maximizes the objective
function is always located inside the interval $\left(0,E_{\Sigma}^{\textrm{R}}-\sum_{i=1,i\neq k}^{N^{\textrm{D}}-1}E_{i}^{\textrm{R}}\right)$,
and the partial derivative of the objective function over each $E_{k}^{\textrm{R}}$
should equal zero, i.e., $\frac{dD{}^{\textrm{R}}\left(T\right)}{dE_{k}^{\textrm{R}}}=0$
for $k=1,2,...,N^{\textrm{D}}-1$. As a result, for the optimal solution,
the following equations must be satisfied 
\begin{align}
T_{k}^{\textrm{S}}g\left(\frac{E{}_{k}^{\textrm{D}}}{T{}_{k}^{\textrm{S}}}\right) & =\left(T_{k}^{\textrm{D}}-T_{k}^{\textrm{S}}\right)g\left(\frac{E_{k}^{\textrm{R}}}{T{}_{k}^{\textrm{D}}-T_{k}^{\textrm{S}}}\right),\label{eq:13}\\
\frac{dD{}^{\textrm{R}}\left(T\right)}{dE_{k}^{\textrm{R}}} & =0,\begin{array}{c}
\end{array}\begin{array}{c}
\end{array}k=1,2,...,N^{\textrm{D}}-1,\label{eq:14}\\
\sum_{k=1}^{N^{\textrm{D}}}E_{k}^{\textrm{R}} & =E_{\Sigma}^{\textrm{R}}.\label{eq:15}
\end{align}
Consequently, the necessity of the equation set is proved.

\noindent On the other hand, both Eqns. (\ref{eq:13}) and (\ref{eq:14})
have a unique solution for each $k$, and we can further verify that
the whole equation set also has a unique solution. Therefore, the
sufficiency of this equation set is also proved.

\section*{Appendix 3 \ \ \ \ \ \ \ Proof of Theorem \ref{thm:All-the-DWF}}

It is obvious that the DWF points obtained through Algorithm \ref{alg:TMIN_power}
satisfy all the other parts of Lemma \ref{lem:DWF points} except
for (\ref{eq:9}). In Algorithm \ref{alg:TMIN_power}, when determining
$i_{k}$, i.e., the index of the $k$-th DWF interval in all the energy
arrival epochs, we have only checked \textit{
\begin{equation}
i_{k}=\arg\underset{i<\tilde{i_{k}}}{\min}\left\{ \frac{\sum_{j=i_{k-1}+1}^{i}E_{j}}{t{}_{i}-t_{i_{k-1}}}\right\} ,\label{eq:16}
\end{equation}
}which is a local operation, but Lemma \ref{lem:DWF points} requires
a global condition 
\begin{equation}
i_{k}=\arg\underset{i:t_{i}\leq T}{\min}\left\{ \frac{\sum_{j=i_{k-1}+1}^{i}E_{j}}{t{}_{i}-t_{i_{k-1}}}\right\} .\label{eq:17}
\end{equation}
Hence, we need to prove that the former expression infers the latter
one, which will be shown by contradiction. For an arbitrary $k$,
we first get $i_{k}$ from (\ref{eq:16}), and $i_{k}'$ from (\ref{eq:17}),
and we assume that $i_{k}'>\tilde{i}_{k}>i_{k}$. Based on the original
EH profile, we can construct a new EH profile, of which the part $t\in\left[0,t_{i_{k}}\right]$
is the same as the original EH profile, while in the interval $\left(t_{i_{k}},t_{i'_{k}}\right]$
there is no energy arrival. Due to the third property of DWF, a constant
source power transmission for both EH profiles should be adopted in
$\left[0,t_{i'_{k}}\right]$. Furthermore, we can verify that the
feasibility of $i'_{k}$ for the new EH profile is equivalent to that
of the original EH profile, the proof of which is omitted. We can
then focus on the proof for the new EH profile. 

\begin{figure}
\begin{centering}
\includegraphics[scale=0.5]{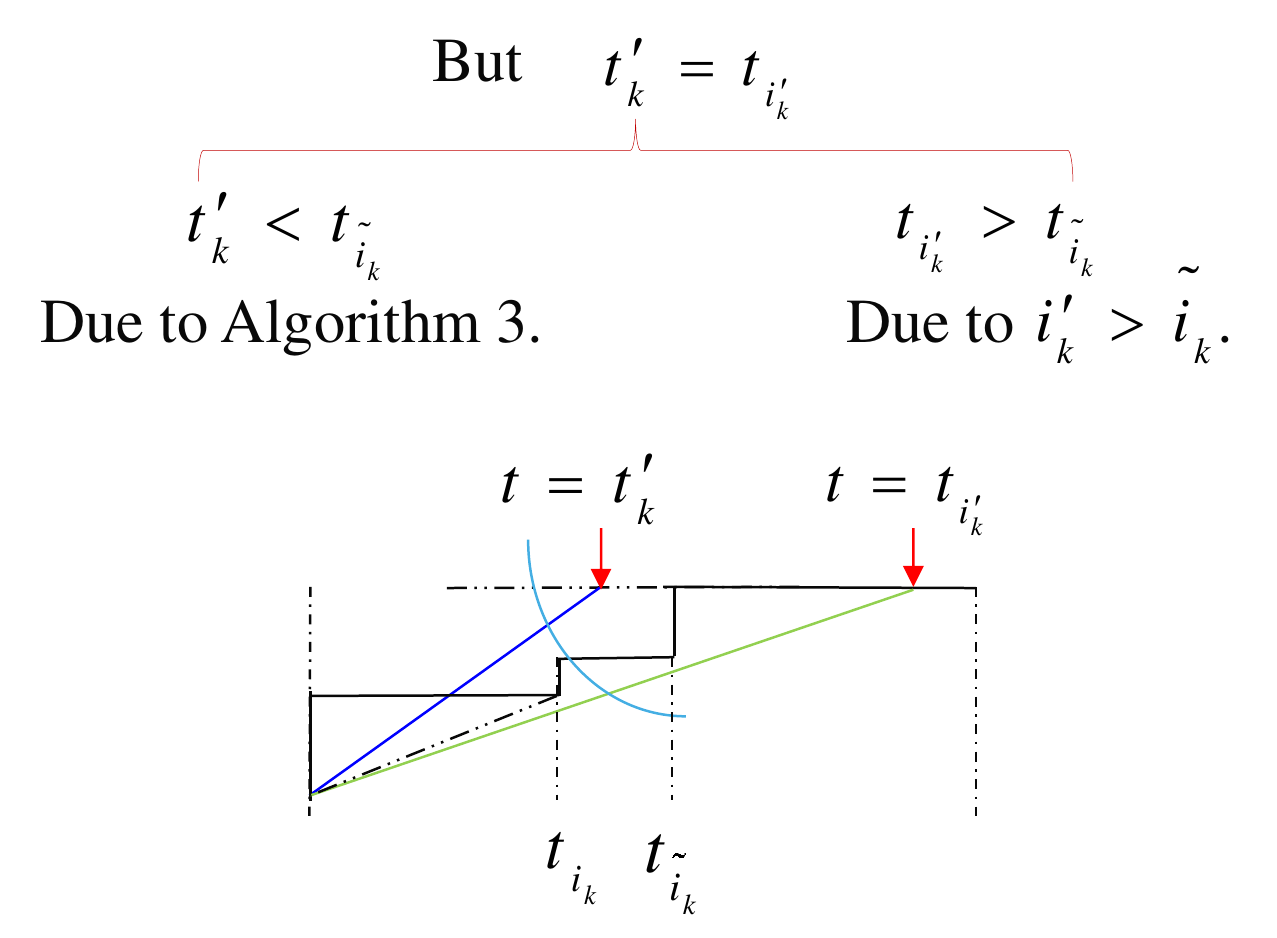}
\par\end{centering}

\caption{\label{fig:The-contradiction-of}The contradiction of time, for the
proof of Theorem \ref{thm:All-the-DWF}.}
\end{figure}

For the new EH profile, since $i'_{k}>\tilde{i_{k}}$, we have $t_{i'_{k}}>t_{\tilde{i_{k}}}$.
On the other hand, due to the calculation procedure in the algorithm,
the obtained $t_{k}'$ should satisfy $t_{k}'<t_{\tilde{i_{k}}}$.
However, based on the fact that the equation $g\left(\frac{E_{0}}{t}\right)t=D_{0}$
over $t$ with arbitrary positive $D_{0}$ and $E_{0}$ has a unique
solution (as long as it has a solution), we can obtain $t_{i'_{k}}=t_{k}'$.
This contradiction of $t_{\tilde{i_{k}}}<t_{i'_{k}}=t_{k}'<t_{\tilde{i_{k}}}$
is illustrated in Fig. \ref{fig:The-contradiction-of}. Therefore,
the assumed case does not hold for the new EH profile, nor for the
original EH profile.

\bibliographystyle{IEEEtran}
\bibliography{example}

\end{document}